\documentclass{aa}  
\usepackage{natbib}
\bibpunct{(}{)}{;}{a}{}{,}
\usepackage{graphicx}
\usepackage{enumitem}
\usepackage{epstopdf}
\usepackage[percent]{overpic}
\usepackage{color}

\begin{document}

\authorrunning{Chatzistergos et al.}
\titlerunning{Plage area composite}
\title{Analysis of full disc Ca~II~K spectroheliograms \\III. Plage area composite series covering 1892--2019}
\newcounter{affiliations}
\author{Theodosios~Chatzistergos\inst{\refstepcounter{affiliations}{\theaffiliations}\label{INAF},\refstepcounter{affiliations}{\theaffiliations}\label{MPS}}, 
	Ilaria~Ermolli\inst{\ref{INAF}},  
	Natalie~A.~Krivova\inst{\ref{MPS}}, 
	Sami~K.~Solanki\inst{\ref{MPS},\addtocounter{affiliations}{1}\theaffiliations}, 
	Dipankar~Banerjee\inst{\refstepcounter{affiliations}{\theaffiliations}\label{IIAP},\refstepcounter{affiliations}{\theaffiliations}\label{ARIES}}, 
	Teresa~Barata\inst{\refstepcounter{affiliations}{\theaffiliations}\label{CESR1}}, 
	Marcel~Belik\inst{\refstepcounter{affiliations}{\theaffiliations}\label{UPICE}}, 
	Ricardo~Gafeira\inst{\ref{CESR1},\addtocounter{affiliations}{1}\theaffiliations}, 
	Adriana~Garcia\inst{\ref{CESR1}}, 
	Yoichiro~Hanaoka\inst{\refstepcounter{affiliations}{\theaffiliations}\label{NAOJ}},
	Manjunath~Hegde\inst{\ref{IIAP}},
	Jan~Klime\v{s}\inst{\ref{UPICE}},
	Viktor~V.~Korokhin\inst{\refstepcounter{affiliations}{\theaffiliations}\label{KHARKIV}},
	Ana~Louren\c{c}o\inst{\ref{CESR1}}, 
	Jean-Marie~Malherbe\inst{\addtocounter{affiliations}{1}\theaffiliations,\addtocounter{affiliations}{1}\theaffiliations},
	Gennady~P.~Marchenko\inst{\ref{KHARKIV}},
	Nuno~Peixinho\inst{\ref{CESR1}},
	Takashi~Sakurai\inst{\ref{NAOJ}},
	Andrey~G.~Tlatov\inst{\addtocounter{affiliations}{1}\theaffiliations}} 
\offprints{Theodosios Chatzistergos  \email{chatzistergos@mps.mpg.de, theodosios.chatzistergos@inaf.it}}

\institute{INAF Osservatorio Astronomico di Roma, Via Frascati 33, 00078 Monte Porzio Catone, Italy 
\and Max Planck Institute for Solar System Research, Justus-von-Liebig-weg 3, 37077 G\"{o}ttingen, Germany 
\and School of Space Research, Kyung Hee University, Yongin, Gyeonggi 446-701, Republic of Korea 
\and Indian Institute of astrophysics, Koramangala, Bangalore 560034, India
\and Aryabhatta Research Institute of Observational Sciences,  Nainital-263 001, Uttarakhand, India
\and Univ Coimbra, CITEUC - Center for Earth and Space Research of the University of Coimbra, Geophysical and Astronomical Observatory, 3040-004 Coimbra, Portugal
\and Observatory Upice, U lipek 160, 542 32 \'Upice, Czech republic
\and Instituto de Astrof\'{i}sica de Andaluc\'{i}a (CSIC), Apartado de Correos 3004, E-18080 Granada, Spain 
\and National Astronomical Observatory of Japan 2-21-1 Osawa, Mitaka, Tokyo 181-8588, Japan
\and Astronomical Institute of Kharkiv V.N. Karazin National University, 35 Sumskaya St., Kharkiv, 61022, Ukraine
\and LESIA, Observatoire de Paris, 92195 Meudon, France
\and PSL Research University, Paris, France 
\and Kislovodsk Mountain Astronomical Station, Central Astronomical Observatory,
Russian Academy of Sciences, Kislovodsk, Russia}
\date{}
        
\abstract
{Studies of long-term solar activity and variability require knowledge of the past evolution of the solar surface magnetism.
An important source of such information are the archives of full-disc Ca~II~K observations performed more or less regularly at various sites since 1892. }
{We derive the plage area evolution over the last 12 solar cycles employing data from all Ca~II~K archives available publicly in digital form known to us, including several as yet unexplored Ca~II~K archives.} 
{We analyse more than 290,000 full-disc Ca~II~K observations from 43 datasets spanning the period 1892--2019. 
All images were consistently processed with an automatic procedure that performs the photometric calibration (if needed) and the limb-darkening compensation.
The processing also accounts for artefacts plaguing many of the images, including some very specific artefacts such as bright arcs found in Kyoto and Yerkes data.
The employed methods have previously been tested and evaluated on synthetic data and found to be more accurate than other methods used in the literature to treat a subset of the data analysed here. } 
{We have produced a plage area time-series from each analysed dataset. 
We found that the differences between the plage areas derived from individual archives are mainly due to the differences in the central wavelength and the bandpass used to acquire the data at the various sites. 
We have empirically cross-calibrated and combined the results obtained from each dataset to produce a composite series of plage areas. 
"Backbone" series are used to bridge all the series together.
We have also shown that the selection of the backbone series has  little effect on the final plage area composite. We have quantified the uncertainty of determining the plage areas with our processing due to shifts in the central wavelength and found it to be less than 0.01 in fraction of the solar disc for the average conditions found on historical data.
We also found the variable seeing conditions during the observations to slightly increase the plage areas during activity maxima.}
{We provide the so far most complete time series of plage areas based on corrected and calibrated historical and modern Ca~II~K images. Consistent plage areas are now available on 88\% of all days from 1892 onwards and on 98\% from 1907 onwards.}

        
\maketitle
        
\section{Introduction}
\label{sec:intro}
\sloppy
\begin{figure*}[t!]   
	\centering 
	\includegraphics[width=1\linewidth]{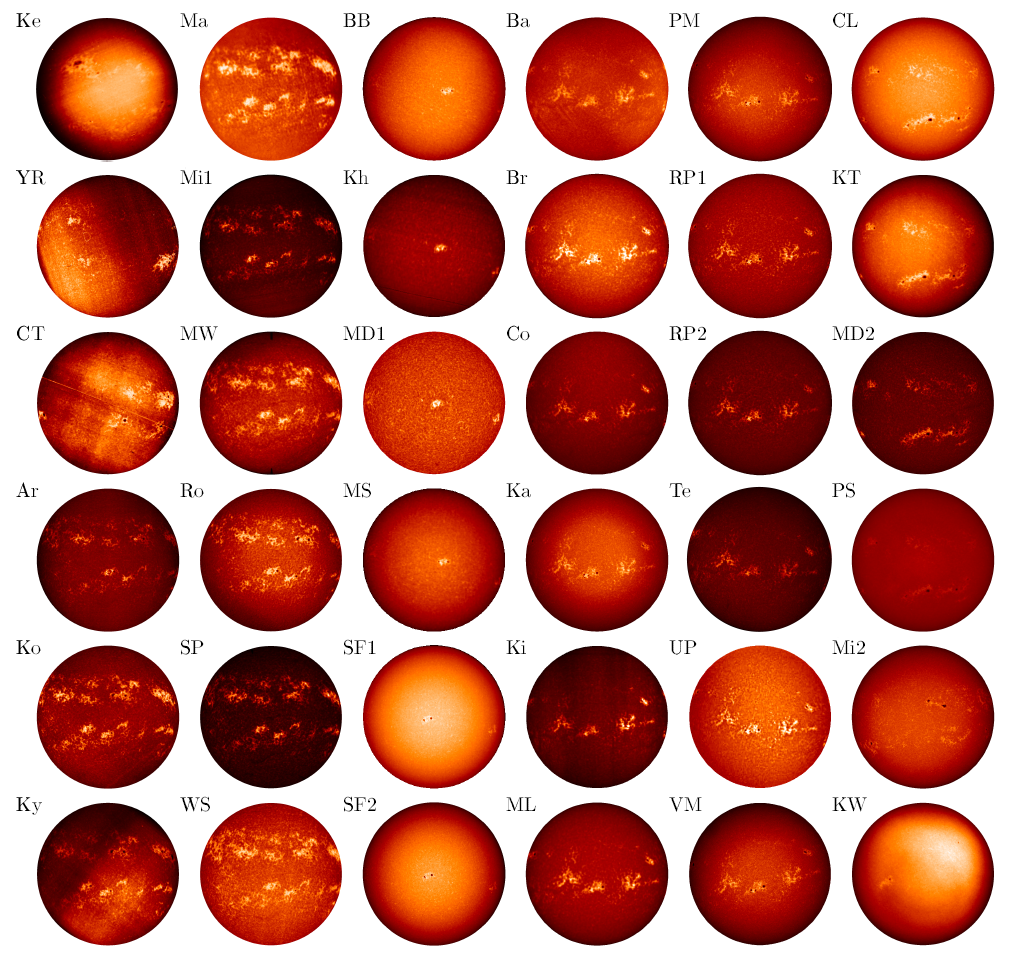}
	\caption{Examples of observations from the various archives analysed in this study, except from MM and Sc. With the exception of the images in the first and last column,  the images within each column correspond to roughly the same day. Within a column the images are shown in alphabetical order according to the name of the observatory, given by a 2-letter abbreviation, with a numeral added in some cases (see Table 1 for the corresponding observatory name). In particular, the dates of the observations are: 14/07/1892 for Ke; 16/02/1907 for YR; 04/09/1908 for CT; 03/02/1968 for Ma; 04/02/1968 for Ko, Ky, Mi1, MW, and SP; 05/02/1968 for Ar, Ro, and WS; 16/07/1995 for BB, Kh, MD1, and MS; 15/07/1995 for SF1 and SF2; 13/03/2014 for Ba, Br, Co, Ka, Ki, ML, PM, RP1, RP2, Te, UP, and VM; 01/08/2012 for CL, KT, MD2, and PS; 10/07/2015 for Mi2;  23/04/2018 for KW; respectively. The images are shown after the preprocessing to identify the disc and re-sample them to account for the disc's ellipticity (when applicable) and convert the historical data to density values. The images have been roughly aligned to show the solar north pole at the top.}
\label{fig:processedimages_raw} 
\end{figure*}                 

There is a need to understand the long-term solar magnetic activity, which is also important for Earth's climate studies \citep[][]{haigh_sun_2007,gray_solar_2010,ermolli_recent_2013,solanki_solar_2013-1}.
This requires long and reliable solar activity indices \citep[e.g.][]{kopp_impact_2016,yeo_empire:_2017,shapiro_nature_2017,wu_solar_2018-2}.

\begin{figure*}
	\centering
	\includegraphics[width=1\linewidth]{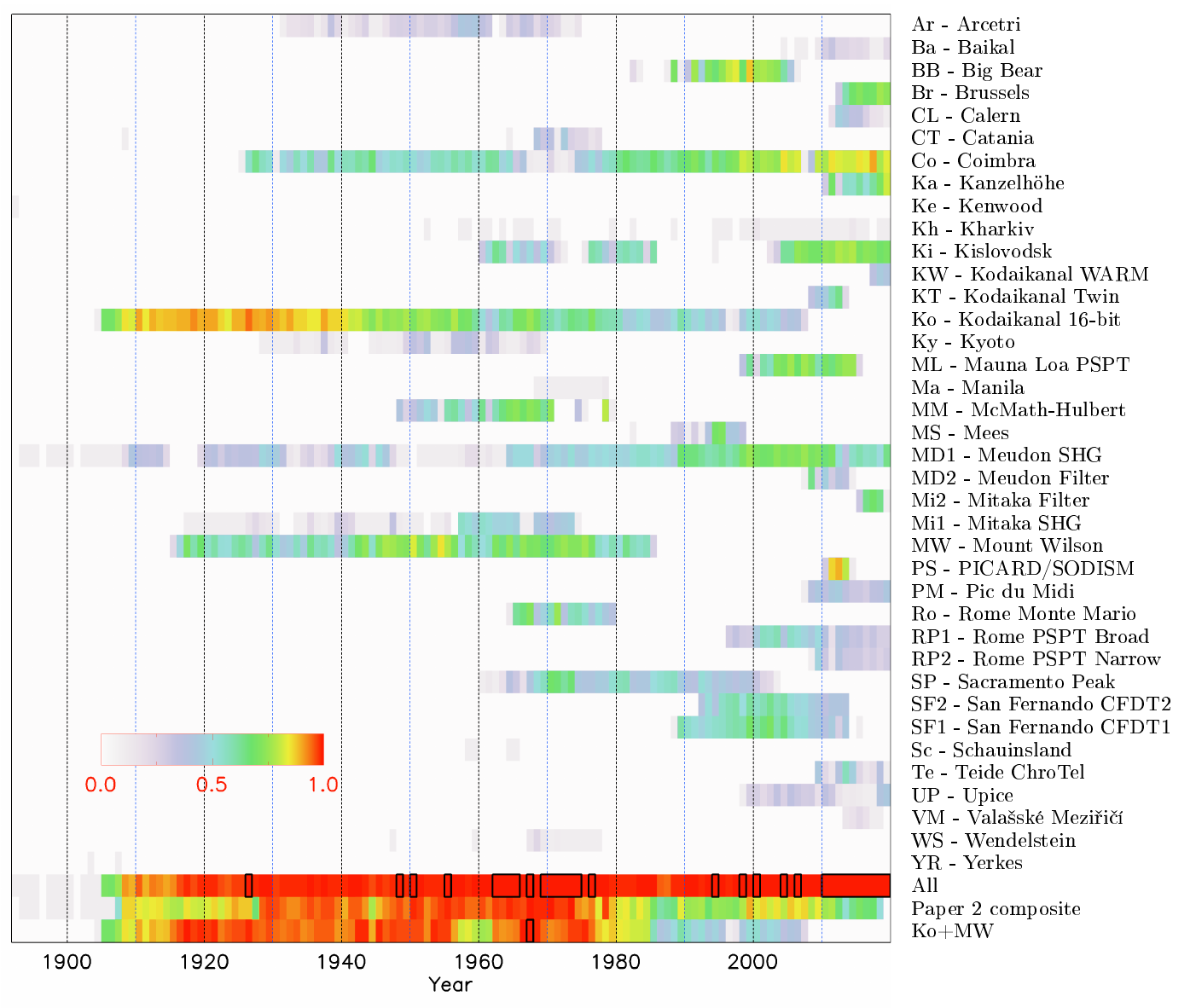}
	\caption{Annual coverage of the various Ca~II~K archives analysed in this study, except from those taken off-band (see Sect. \ref{sec:data} for details). Also shown is the annual coverage of all the archives combined, the one for the \citet[][paper 2]{chatzistergos_analysis_2019} composite series, and the annual coverage when only Ko and MW are considered. The annual coverage is colour-coded as shown by the colour bar plotted in the bottom left corner of the panel. Years with full coverage are outlined with black rectangles.}
	\label{fig:timeline3}
\end{figure*}

One such dataset comes from the collection of full-disc Ca~II~K observations.
Numerous Ca~II~K archives, recorded and stored at various observatories all over the world, were digitised over the last 3 decades.
Whether in physical or digital form, such archives have been used to derive information about the evolution of plage regions
\citep[e.g.][and references therein]{antonucci_chromospheric_1977,ermolli_comparison_2009,dorotovic_north-south_2010,chatzistergos_exploiting_2016,chatzistergos_analysis_2019,barata_software_2018,tlatov_polar_2019}, the solar radius variations \citep{meftah_solar_2018,hiremath_nearly_2020}, network cell properties \citep[e.g.][]{berrilli_average_1999,ermolli_measure_2003,chatterjee_variation_2017,raju_temporal_2018}, photometric properties of disc features over the solar cycle \citep[e.g.][]{ermolli_photometric_2007,ermolli_analysis_2010}, as well as for reconstructions of irradiance variations \citep[e.g.][]{ermolli_recent_2011,chapman_comparison_2012,fontenla_bright_2018} and studies of the relation between the Ca II K brightness and the photospheric magnetic field \citep[e.g.][and references therein]{loukitcheva_relationship_2009,chatzistergos_recovering_2019}.

The results presented in the literature on the plage area evolution show considerable discrepancies \citep[][]{chatzistergos_analysis_2017,chatzistergos_analysis_2019,ermolli_potential_2018}. 
Indeed, a critical aspect of such studies is that the accuracy of the processing applied to the data has not been evaluated.
Also, only few individual archives have been used until now, and the results from various archives have not been combined.
This is largely because the techniques used for the data analysis were specifically developed for each single archive and could not be directly applied to different data. 
The most employed archives are those from the Arcetri \citep{ermolli_digitized_2009}, Kodaikanal \citep{chatterjee_butterfly_2016}, Mt Wilson \citep{lefebvre_solar_2005}, and Sacramento Peak \citep{tlatov_new_2009} observatories.

\newcounter{tableid}
\begin{table*}
	\caption{List of Ca~II~K datasets analysed in this study.}              
	\label{tab:observatories}      
	\centering                                      
	\small
	\begin{tabular}{l*{8}{c}}          
		\hline\hline                        
		Observatory & Acronym	   &Detector &Instrument&Period		  &Images&SW   			          &Pixel scale			  	  & Ref.\\
		&   		   &		 &			& 		      &      &[$\text{\AA}$]		  &[$"/$pixel] 			  	  &	\\
		\hline
		Arcetri\tablefootmark{a} &Ar		   &Plate	 & SHG		& 1931--1974  & 4871 &0.3   				  &2.5	  			          &\addtocounter{tableid}{1}\thetableid\\  
		Baikal				 &Ba		   &CCD	 	 & Filter	& 2004--2019  & 846  &1.2					  &2.7 				 	  &\addtocounter{tableid}{1}\thetableid\\
		Big Bear  	 		 &BB		   &CCD	 	 & Filter 	& 1982--2006  & 5027 &3.2, 1.5\tablefootmark{b}&4.2, 2.4\tablefootmark{c} &\addtocounter{tableid}{1}\thetableid\\ 
		Brussels   			 &Br		   &CCD	 	 & Filter 	& 2012--2019  &14699 &2.7					  & 1.0     				  &\addtocounter{tableid}{1}\thetableid\\   
		Calern			     &CL		   &CCD	     & Filter	& 2011--2019  & 1560	 &7						  & 1.0						  &\addtocounter{tableid}{1}\thetableid\\
		Catania				 &CT		   &Plate    & SHG      & 1908--1977  & 1008 &- 		   			  &1.1--5					  &\addtocounter{tableid}{1}\thetableid\\
		Coimbra	   			 &Co&Plate/CCD\tablefootmark{d}& SHG& 1925--2019  & 19758 &0.16      			  & 2.2 					  &\addtocounter{tableid}{1}\thetableid\\
		Kanzelh\"ohe		 &Ka		   &CCD	 	 & Filter   & 2010--2019  & 8550 &3.0					  &	1.0 					  &\addtocounter{tableid}{1}\thetableid\\
		Kenwood				 &Ke		   &Plate	 & SHG	    & 1892  	  & 5    &-						  &	3.1						  &\addtocounter{tableid}{1}\thetableid\\
		Kharkiv			 &Kh&Plate/CCD\tablefootmark{e}&SHG & 1952--2019  & 564  &3.0					  &	3.3						  &\addtocounter{tableid}{1}\thetableid\\  
		Kislovodsk			 &Ki		   &Plate/CCD\tablefootmark{f}& SHG&1960--2019&9738 &-  			  &1.3, 2.3\tablefootmark{f}						  &\addtocounter{tableid}{1}\thetableid\\
		Kodaikanal\tablefootmark{a}			 &Ko		   &Plate	 & SHG 		& 1904--2007  & 45047&0.5 				      & 0.9	  				  	  &\addtocounter{tableid}{1}\thetableid\\ 
		Kodaikanal Twin		 &KT		   &CCD 	 & Filter	& 2008--2013  & 3059 &1.2 				      & 1.2	  				  	  &\addtocounter{tableid}{1}\thetableid\\ 
		Kodaikanal WARM		 &KW		   &CCD 	 & Filter	& 2017--2019  & 585  & 1.0 				      & 2.4	  				  	  &\addtocounter{tableid}{1}\thetableid\\ 
		Kyoto				 &Ky		   &Plate    & SHG	    & 1928--1969  & 3119 &0.74					  &	2.0						  &\addtocounter{tableid}{1}\thetableid\\ 
		Manila				 &Ma		   &Plate	 & SHG		& 1968--1978  &	162	 &0.5					  &	1.2						  &\addtocounter{tableid}{1}\thetableid\\ 	
		Mauna Loa PSPT  	 &ML		   &CCD	 	 & Filter 	& 1998--2015  &31933 &2.7 					  & 1.0	  					  &\addtocounter{tableid}{1}\thetableid\\
		McMath-Hulbert\tablefootmark{a}		 &MM		   &Plate	 & SHG		& 1948--1979  & 4932 &0.1 				      & 3.1        			  	  &\addtocounter{tableid}{1}\thetableid\\
		Mees	   			 &MS		   &CCD	 	 & Filter 	& 1982--1998  & 1519 &1.2 					  & 5.5						  &\addtocounter{tableid}{1}\thetableid\\
		Meudon\tablefootmark{a}	   			 &MD1 &Plate/CCD\tablefootmark{g}&SHG& 1893--2019 &20117 &0.15, 0.09\tablefootmark{h}					  & 2.2, 1.5, 1.1\tablefootmark{i}			  &\addtocounter{tableid}{1}\thetableid\\ 
		Meudon	   			 &MD2 		   &CCD		 &Filter	& 2007--2014  &1519  &1.4					  & 0.9						  &\addtocounter{tableid}{1}\thetableid\\ 
		Mitaka\tablefootmark{a}  			 &Mi1		   &Plate	 & SHG	    & 1917--1974  & 4193 &0.5 					  &0.9, 0.7\tablefootmark{j}  &\addtocounter{tableid}{1}\thetableid\\
		Mitaka  			 &Mi2		   &CCD		 & Filter   & 2015--2019  & 897  &4.5 					  &1.0						  &\addtocounter{tableid}{1}\thetableid\\
		Mount Wilson\tablefootmark{a} 		 &MW		   &Plate	 & SHG	    & 1915--1985  & 39545&0.2 					  &2.9	  				  	  &\addtocounter{tableid}{1}\thetableid\\
		Pic du Midi			 &PM		   &CCD	 	 & Filter 	& 2007--2019  & 3794 &2.5 					  & 1.2 	  				  &\addtocounter{tableid}{1}\thetableid\\
		PICARD/SODISM		 &PS		   &CCD		 & Filter	& 2010--2014  &	1218  &7						  &1						  &\addtocounter{tableid}{1}\thetableid\\
		Rome Monte Mario	 &Ro 		   &Plate 	 & Filter 	& 1964--1979  & 5826 &0.3 					  & 5.0						  &\addtocounter{tableid}{1}\thetableid\\
		Rome PSPT\tablefootmark{a}		     &RP1   	   &CCD		 & Filter	& 1996--2019  & 3449 &2.5 					  &2.0\tablefootmark{k}	  	  &\addtocounter{tableid}{1}\thetableid\\
		Rome PSPT		     &RP2		   &CCD		 & Filter	& 2008--2019  & 1298 &1.0 					  &2.0\tablefootmark{k}	  	  &\thetableid\\ 
		Sacramento Peak 	 &SP		   &Plate 	 & SHG  	& 1960--2002  & 7750 &0.5  					  &1.2						  &\addtocounter{tableid}{1}\thetableid\\
		San Fernando CFDT1	 &SF1		   &CCD	 	 & Filter 	& 1988--2015  & 4986 &9  					  &5.1						  &\addtocounter{tableid}{1}\thetableid\\
		San Fernando CFDT2	 &SF2  		   &CCD	 	 & Filter 	& 1992--2013  & 4065 &9  					  &2.6 						  &\thetableid\\ 
		Schauinsland\tablefootmark{a}	     &Sc 		   &Plate	 & SHG		& 1958--1965  &18    &-   					  &1.7, 2.6\tablefootmark{l}  &\addtocounter{tableid}{1}\thetableid\\
		Teide ChroTel		 &Te		   &CCD	 	 & Filter 	& 2009--2019  & 1843 &0.3 					  &1.0 					 	  &\addtocounter{tableid}{1}\thetableid \\  
		Upice				 &UP		   &CCD		 & Filter	& 1998--2019  &	3234 &1.6					  &4.0, 2.4\tablefootmark{m}  &\addtocounter{tableid}{1}\thetableid \\  
		Vala\v{s}sk\'{e} Mezi\v{r}i\v{c}\'{i} &VM &CCD & Filter & 2011--2018  & 318  &2.4 			  		  &1.8				  		  &\addtocounter{tableid}{1}\thetableid \\ 
		Wendelstein\tablefootmark{a}  		 &WS 		   &Plate	 & SHG		& 1947--1977  &422   &-   					  &1.7, 2.6\tablefootmark{l}  &\addtocounter{tableid}{-3}\thetableid\addtocounter{tableid}{3}\\
		Yerkes				 &YR		   &Plate	 & SHG		& 1903--1907  &7	 &-						  &2.4						  &\addtocounter{tableid}{1}\thetableid \\ 
		\hline
	\end{tabular}
	\tablefoot{Columns are: name of the observatory, abbreviation used in this study, type of detector, type of instrument, period of observations, total number of images (including multiple images on a single day when available) analysed in this study, spectral width of the spectrograph/filter, average pixel scale of the images, and the bibliography entry. \tablefoottext{a}{These archives were considered in Paper 2, although in the case of the Ko data the earlier 8-bit digitisation was used.} \tablefoottext{b}{The two values correspond to the period before and after 10/09/1996.} \tablefoottext{c}{The two values correspond to the period before and after 08/11/1995, when the CCD camera was upgraded.}\tablefoottext{d}{The CCD camera was installed in January 2007, but observations were stored on photographic plates up to December 2007.} \tablefoottext{e}{The CCD camera was installed on 01/09/1994.} \tablefoottext{f}{The CCD camera was installed on 18/12/2002.} \tablefoottext{g}{The observations were stored on photographic plates up to 27/09/2002, while observations with a CCD camera started on 13/05/2002.} \tablefoottext{h}{The values refer to the periods before and after 15/06/2017.} \tablefoottext{i}{The values refer to the periods [24/10/1893--27/09/2002], [28/09/2002--14/06/2017], and since 15/06/2017.} \tablefoottext{j}{These data derive from two digitisations and the two values correspond to the earlier and more recent digitisation, respectively. See \cite{chatzistergos_analysis_2019} for more information.} 
		\tablefoottext{k}{The pixel scale is for the resized images to match the seeing conditions of the observing location.}  
		\tablefoottext{l}{These data were stored in TIFF and JPG files with different spatial resolution, the values correspond to the TIFF and JPG files, respectively.}
		\tablefoottext{m}{The two values correspond to the period before and after 01/01/2018, when the CCD camera was upgraded.}}
	\tablebib{\addtocounter{tableid}{-\thetableid}
		(\addtocounter{tableid}{1}\thetableid) \citet{ermolli_digitized_2009}; 
		(\addtocounter{tableid}{1}\thetableid) \citet{golovko_data_2002}; 
		(\addtocounter{tableid}{1}\thetableid) \citet{naqvi_big_2010}; 
		(\addtocounter{tableid}{1}\thetableid) \url{http://www.sidc.be/uset/}; 
		(\addtocounter{tableid}{1}\thetableid) \citet{meftah_solar_2018};
		(\addtocounter{tableid}{1}\thetableid) \citet{zuccarello_solar_2011};
		(\addtocounter{tableid}{1}\thetableid) \citet{garcia_synoptic_2011};
		(\addtocounter{tableid}{1}\thetableid) \citet{hirtenfellner-polanec_implementation_2011};
		(\addtocounter{tableid}{1}\thetableid) \citet{hale_solar_1893};
		(\addtocounter{tableid}{1}\thetableid) \citet{belkina_ccd_1996};
		(\addtocounter{tableid}{1}\thetableid) \citet{tlatov_synoptic_2015};
		(\addtocounter{tableid}{1}\thetableid) \citet{priyal_long_2014};
		(\addtocounter{tableid}{1}\thetableid) \citet{singh_twin_2012};
		(\addtocounter{tableid}{1}\thetableid) \citet{pruthvi_two-channel_2015};
		(\addtocounter{tableid}{1}\thetableid) \citet{kitai_digital_2013};
		(\addtocounter{tableid}{1}\thetableid) \citet{miller_new_1965};
		(\addtocounter{tableid}{1}\thetableid) \citet{rast_latitudinal_2008};  
		(\addtocounter{tableid}{1}\thetableid) \citet{mohler_mcmath-hulbert_1968};  
		(\addtocounter{tableid}{1}\thetableid) \url{http://kopiko.ifa.hawaii.edu/KLine/index.shtml};
		(\addtocounter{tableid}{1}\thetableid) \citet{malherbe_new_2019}; 
		(\addtocounter{tableid}{1}\thetableid) \url{http://bass2000.obspm.fr/data_guide.php};
		(\addtocounter{tableid}{1}\thetableid) \citet{hanaoka_long-term_2013};
		(\addtocounter{tableid}{1}\thetableid) \citet{hanaoka_past_2016};
		(\addtocounter{tableid}{1}\thetableid) \citet{lefebvre_solar_2005};
		(\addtocounter{tableid}{1}\thetableid) \citet{koechlin_solar_2019};   
		(\addtocounter{tableid}{1}\thetableid) \citet{meftah_picard_2014};   
		(\addtocounter{tableid}{1}\thetableid) \citet{chatzistergos_historical_2019};
		(\addtocounter{tableid}{1}\thetableid) \citet{ermolli_photometric_2007};
		(\addtocounter{tableid}{1}\thetableid) \citet{tlatov_new_2009};
		(\addtocounter{tableid}{1}\thetableid) \citet{chapman_solar_1997};
		(\addtocounter{tableid}{1}\thetableid) \citet{wohl_old_2005};
		(\addtocounter{tableid}{1}\thetableid) \citet{bethge_chromospheric_2011};
		(\addtocounter{tableid}{1}\thetableid) \citet{klimes_simultaneous_1999}; 				
		(\addtocounter{tableid}{1}\thetableid) \cite{lenza_system_2014}; 
		(\addtocounter{tableid}{1}\thetableid) \citet{hale_rumford_1903}.
	}
\end{table*}

\newcounter{tableidoff}
\begin{table*}
	\caption{List of off-band Ca~II~K datasets analysed in this study.}              
	\label{tab:observatories2}      
	\centering                                      
	\begin{tabular}{l*{6}{c}}          
		\hline\hline                        
		Observatory & Acronym	   &Instrument&Central wavelength&Period		  &Images&SW   			         \\
		&   		   &		 &[$\text{\AA}$]& 		      &      &[$\text{\AA}$]		 	\\
		\hline
		Coimbra\tablefootmark{a}&CoW	   & SHG    	& 3932.3	  &2008--2018  &3113 &0.16	\\ 
		Mauna Loa PSPT  	 &MLW		   & Filter 	& 3936.3	  &2004--2015  &9552 &1.0	\\
		Meudon	   			 &MDV		   & SHG	    & 3933.4	  &2002--2017  &5717 &0.15	\\ 
		Meudon	   			 &MDR		   & SHG	    & 3934.0	  &2002--2017  &5652 &0.15	\\ 
		Meudon\tablefootmark{a}&MDW		   & SHG	    & 3932.3	  &2002--2018  &4632 &0.15	\\ 
		\hline
	\end{tabular}
	\tablefoot{Columns are: name of the observatory, abbreviation used in this study, type of instrument, central wavelength, period of observations, number of images, and the spectral width of the spectrograph/filter. \tablefoottext{a}{Here we restricted our analysis only to the CCD-based data centred at the wing of the line from the CoW and MDW series. Note, however, that offband data from these sources extend back to 1925 and 1893, respectively.} }
\end{table*}

To overcome these limitations, in our previous paper \cite[][Paper 1, hereafter]{chatzistergos_analysis_2018} we introduced a novel approach to process the historical and modern Ca~II~K observations, to  perform  their photometric calibration, to compensate for the intensity centre-to-limb variation (CLV, hereafter), and to account for various artefacts.
By using synthetic data, we also showed that our method can perform the photometric calibration and account for image artefacts with higher accuracy than other methods presented in the literature. 
More importantly, we showed that, as long as the archives are consistent with each other, e.g., centred at the same wavelength and employing the same bandwidth for the observations, the method can be used to derive accurate information on the evolution of plage areas without the need of any adjustments in the processing of the various archives \citep[][Paper 2, hereafter]{chatzistergos_analysis_2019}.
In Paper 2 we applied our method to 85,972 images from 9 Ca~II~K archives to derive plage areas and produce the first composite of plage areas over the entire 20th century.
In particular, we analysed the Ca~II~K archives from the Arcetri, Kodaikanal (8-bit digitisation), McMath-Hulbert, Meudon, Mitaka, Mt Wilson, Rome/PSPT, Schauinsland, and Wendelstein sites.
Five out of the 9 analysed archives were amongst the most studied and prominent ones, while the remaining archives were from less studied data sources.
There are, however, many other Ca~II~K archives that are available and still remain largely unexplored. 
These archives harbour the potential to fill gaps in the available plage series as well as to address inconsistencies among the various archives and within individual archives (e.g. change in data quality, or in the measuring instrument with time).
Moreover, since the work presented in Paper 2, more data from various historical and modern archives became available in digital form. 
In particular, historical data that have been made available in the meantime include those from the latest 16-bit digitisation of the Kodaikanal archive, Catania, Coimbra, Kenwood, Kharkiv, Kyoto, Manila, Rome, Sacramento Peak, and Yerkes observatories, as well as additional data from the Meudon and Mt Wilson archives. 
In this light, we present here results from the up-to-now most comprehensive analysis of historical and modern Ca~II~K observations taken between 1892 and 2019 from 43 different datasets to produce a composite plage area series.

The paper is organised as follows. 
In Section \ref{sec:data} we present the data analysed in our study and the methods applied on the data.  
Our results for the plage areas from individual archives as well as the composite series are presented in Section \ref{sec:results}. 
In Section \ref{sec:discussion} we discuss our results. 
In Section \ref{sec:conclusions} we summarise the results of our study and draw our conclusions.

\begin{figure}
	\centering
	\includegraphics[width=1\linewidth]{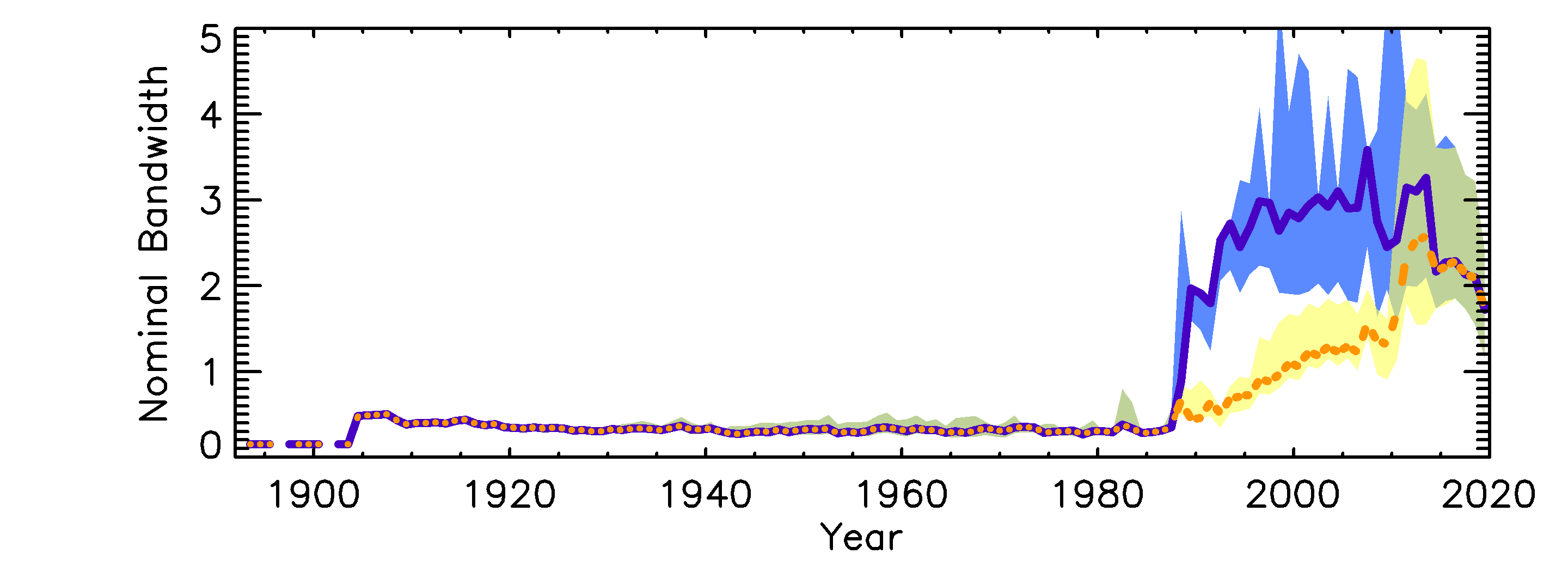}
	\caption{Average nominal bandwidth from all 38 archives included in this study (solid purple) as well as from all archives excluding SF1 and SF2 (dashed orange). The shaded areas show the 1$\sigma$ uncertainty.}
	\label{fig:nominalbandwidths}
\end{figure}

\begin{figure}
	\centering
	\includegraphics[width=1\linewidth]{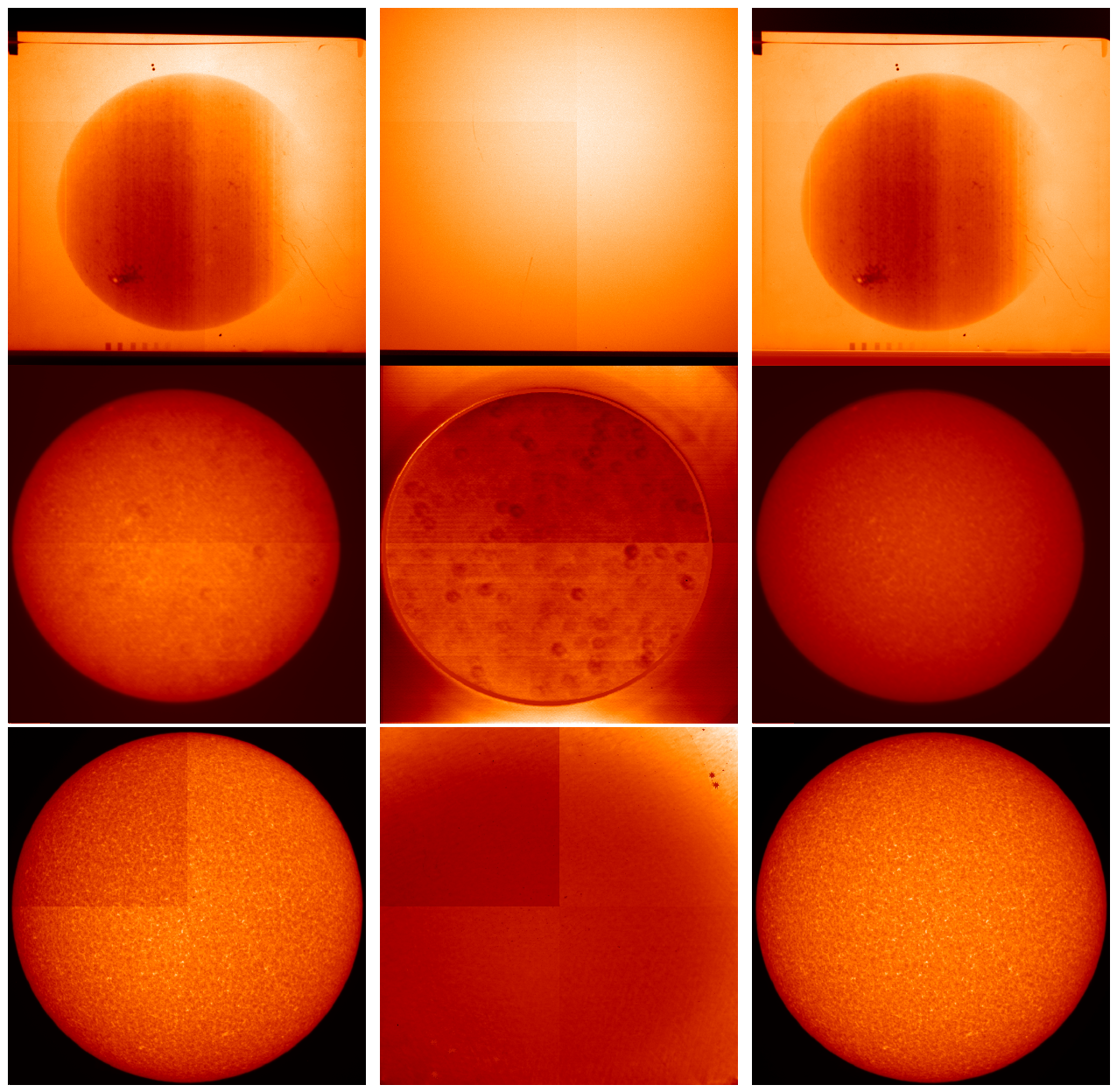}
	\caption{Examples of raw images (left column), data employed for the calibration of the CCD recording device (middle column), and calibrated images (right panels) from the Ko (first row), MS (middle row), and RP1 (bottom row) archives taken on 19/08/2006, 19/12/1995, and 28/02/2020, respectively. The images have not been compensated for ephemeris and are shown in their entire range of values. } 
	\label{fig:examplesflat}
\end{figure}

\section{Data and methods}
\label{sec:data}

\subsection{Ca~II~K observations}
We analysed solar full-disc Ca~II~K observations from 43 datasets, which include series of photographic images and data series acquired with CCD cameras. 
Table \ref{tab:observatories} summarises information on the datasets analysed in our study and their main characteristics.
For the sake of clarity, we also list here all analysed datasets and the corresponding abbreviations used in the text to refer to the various series: Arcetri (Ar), Baikal (Ba), Brussels (Br), Calern (CL), Catania (CT), Coimbra (Co), Kanzelh\"ohe (Ka), Kenwood (Ke), Kharkiv (Kh), Kislovodsk (Ki), Kodaikanal taken with the spectroheliograph (Ko), Kodaikanal taken with the Twin telescope (KT), Kodaikanal taken with the White-Light Active Region Monitor (WARM) telescope (KW), Kyoto (Ky), Manila (Ma), Mauna Loa (ML) taken with the Precision Solar Photometric Telescope (PSPT), McMath-Hulbert (MM), Mees (MS), Meudon taken with the spectroheliograph (MD1), Meudon taken with an interference filter (MD2), Mitaka taken with the spectroheliograph (Mi1), Mitaka taken with the Solar Flare Telescope with an interference filter (Mi2), Mt Wilson (MW), Pic du Midi (PM), SOlar Diameter Imager and Surface Mapper
(SODISM) telescope on board the PICARD spacecraft (PS), Rome taken with the equatorial bar at Monte Mario (Ro), Rome taken with the PSPT (RP1), Rome taken with the PSPT with narrow bandwidth (RP2), Sacramento Peak (SP), San Fernando taken with the Cartesian Full-Disk Telescope (CFDT) 1 (SF1), San Fernando taken with the CFDT2 (SF2), Schauinsland (Sc), Teide (Te) taken with the Chromospheric Telescope (ChroTel), Upice (UP), Vala\v{s}sk\'{e} Mezi\v{r}i\v{c}\'{i} (VM), Wendelstein (WS), and Yerkes (YR).
Figure \ref{fig:processedimages_raw} shows examples of observations from all datasets except for MM and Sc, examples of which can be found in \cite{chatzistergos_ca_2018,chatzistergos_analysis_2019}.
In addition to the 38 datasets listed in Table \ref{tab:observatories}, we also analysed the 5 datasets included in Table \ref{tab:observatories2}.
These 5 datasets include observations centred at different locations of the wing of the Ca~II~K line.

The majority of the analysed datasets stem from observatories located in Europe.
However, there are datasets from Asia and North America that provide an overall good temporal coverage.
All in all, there are 290,147 images taken between 25/06/1892 and 31/12/2019 covering 41,163 days.
Figure \ref{fig:timeline3} shows the fraction of days within a year with at least one observation among all datasets considered in our study.
We also show the total annual coverage by the sum of all the datasets analysed here, the coverage for the case when only Ko and MW are used as well as for the plage area composite presented in Paper 2.
Figure \ref{fig:timeline3} reveals that the data analysed in our study offer a nearly complete coverage, with the exception of the period before 1925, 1944--1946, and 1986--1987.
The coverage is on average 88\% for the whole period of time since 1892. However, it is on average 98\% and is above 76\% for all years if only the period after 1907 is considered.
In contrast, the annual coverage when only Ko and MW are considered is on average 80\%, while it drops  down to 5\% in the 1990's.
The composite provides full annual coverage since 2010 and for 21 more years.
In contrast, the Ko and MW series together provide a full coverage only for 1967.
This illustrates the huge benefit of using multiple datasets to achieve a better coverage over the entire 20th century, but also the need to recover more historical data.
The missing data from Abastumani \citep{khetsuriani_solar_1981}, Anacapri \citep{antonucci_chromospheric_1977}, Baikal, Cambridge \citep{moss_report_1942}, Catania, Crimea, Ebro \citep{curto_historical_2016}, Huairu \citep{suo_full-disk_2020}, Kandilli \citep{dizer_kandilli_1968}, Kenwood, Kharkiv, Kislovodsk, Locarno \citep{waldmeier_swiss_1968}, Madrid \citep{vaquero_spectroheliographic_2007}, Manila, Meudon, Yerkes, Wendelstein, and Schauinsland would be invaluable for this purpose.

\begin{figure*}[t!]   
	\centering 
	\includegraphics[width=1\linewidth]{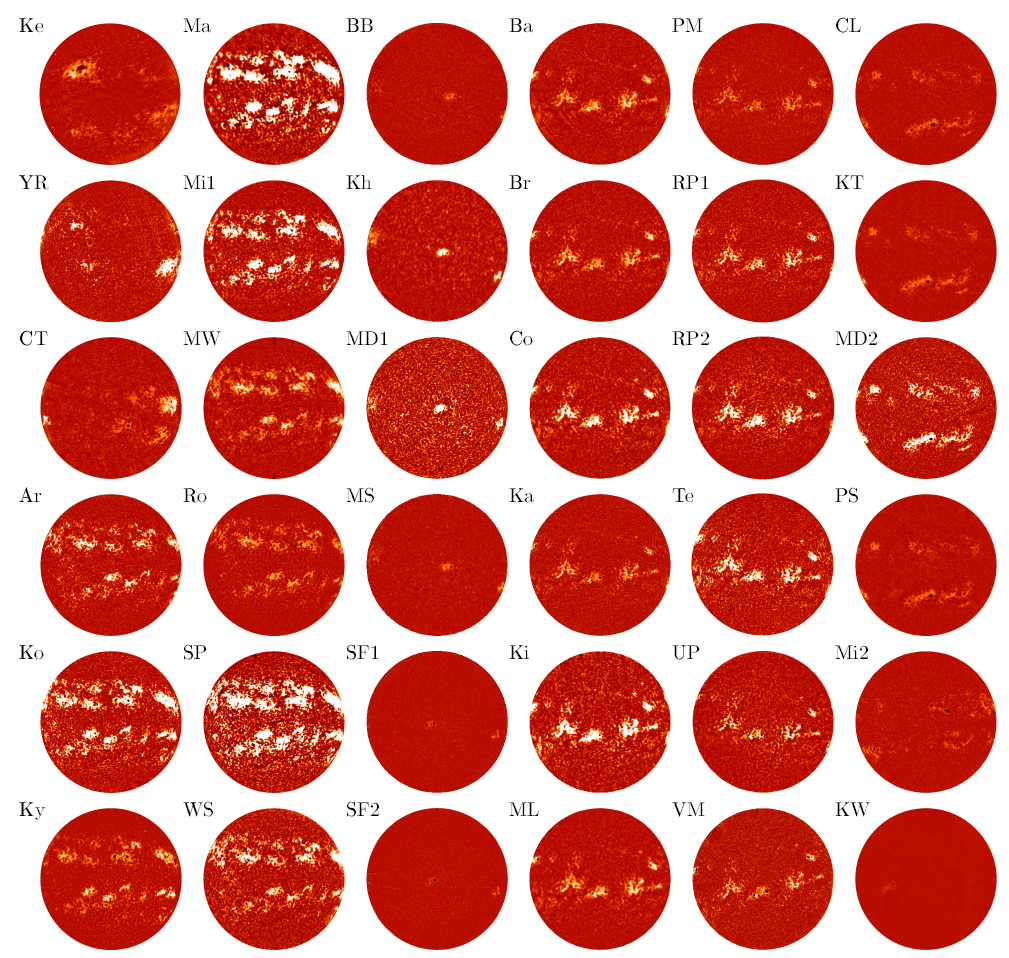}
	\caption{Calibrated and CLV-compensated contrast images of the observations shown in Fig. \ref{fig:processedimages_raw}. Plotted are contrast values in the fixed range of [-0.5,0.5] for all images.}
	\label{fig:processedimages_flat}                            
\end{figure*}  
\begin{figure*}[t!]   
	\centering 
	\includegraphics[width=1\linewidth]{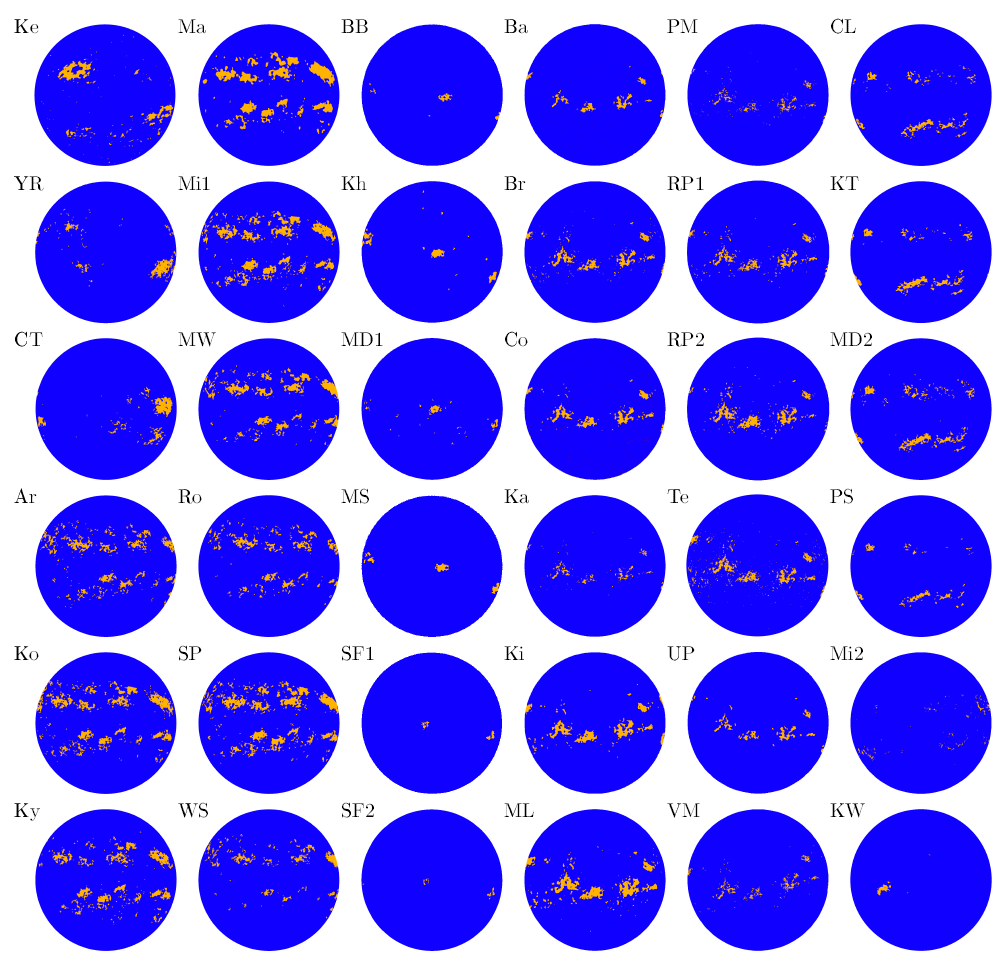}
	\caption{Segmentation masks derived from the observations shown in  Figs. \ref{fig:processedimages_raw} and \ref{fig:processedimages_flat}. Plage are highlighted in yellow, while quiet Sun and network regions together form the blue background. We stress that the same threshold was used for all datasets to identify the plage regions, which is why the different datasets seem to give rather different plage coverage depending on the employed bandwidth or central wavelength.}
	\label{fig:processedimages_mask} 
\end{figure*}        

\begin{figure*}
	\centering
	\includegraphics[width=0.95\linewidth]{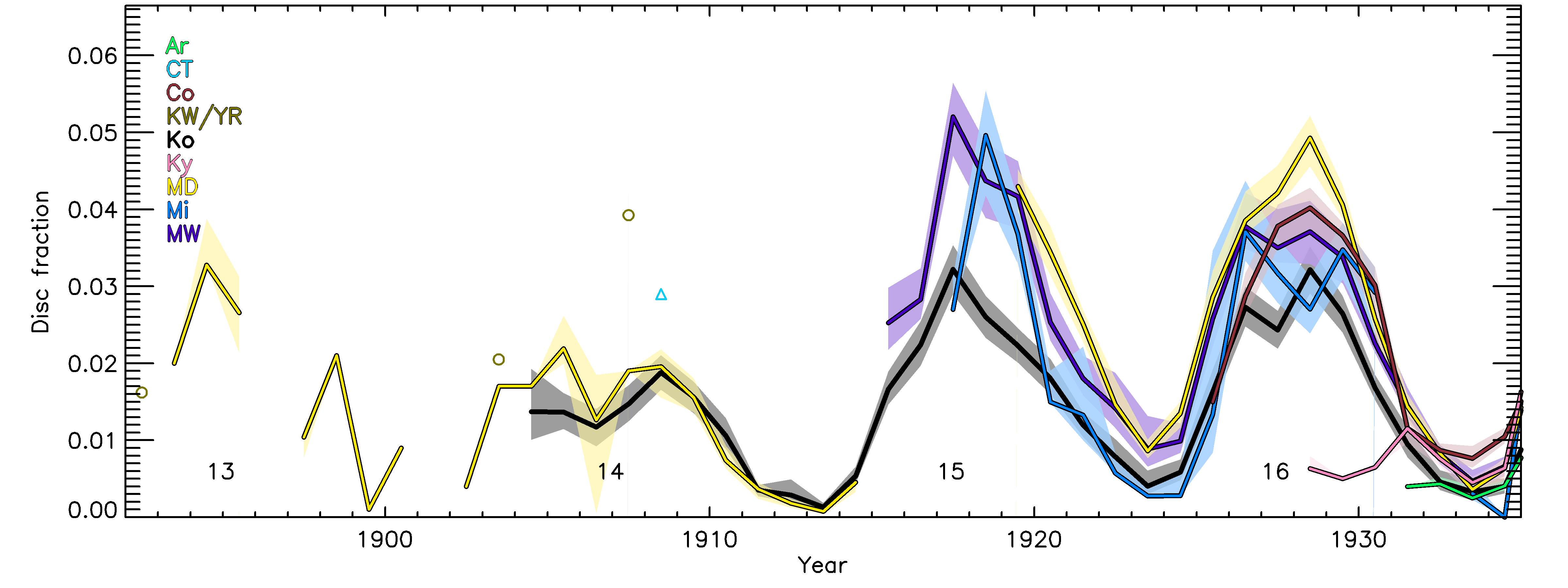}
	\includegraphics[width=0.95\linewidth]{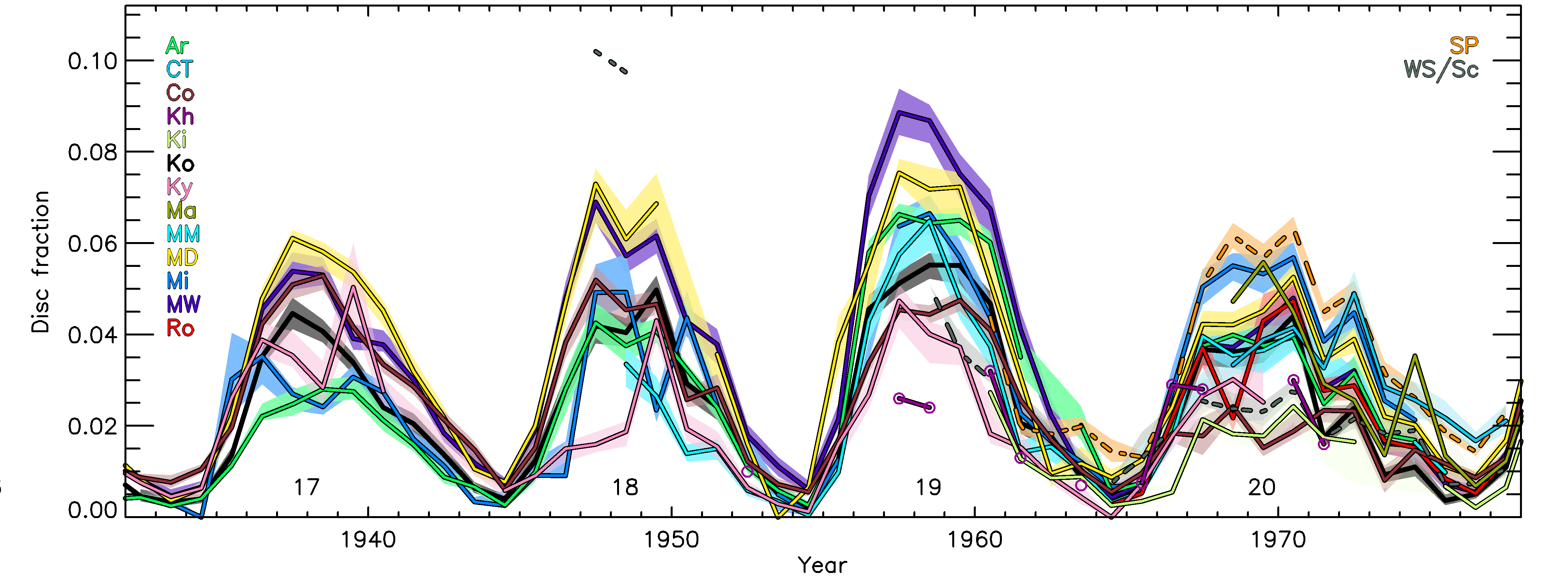}
	\includegraphics[width=0.95\linewidth]{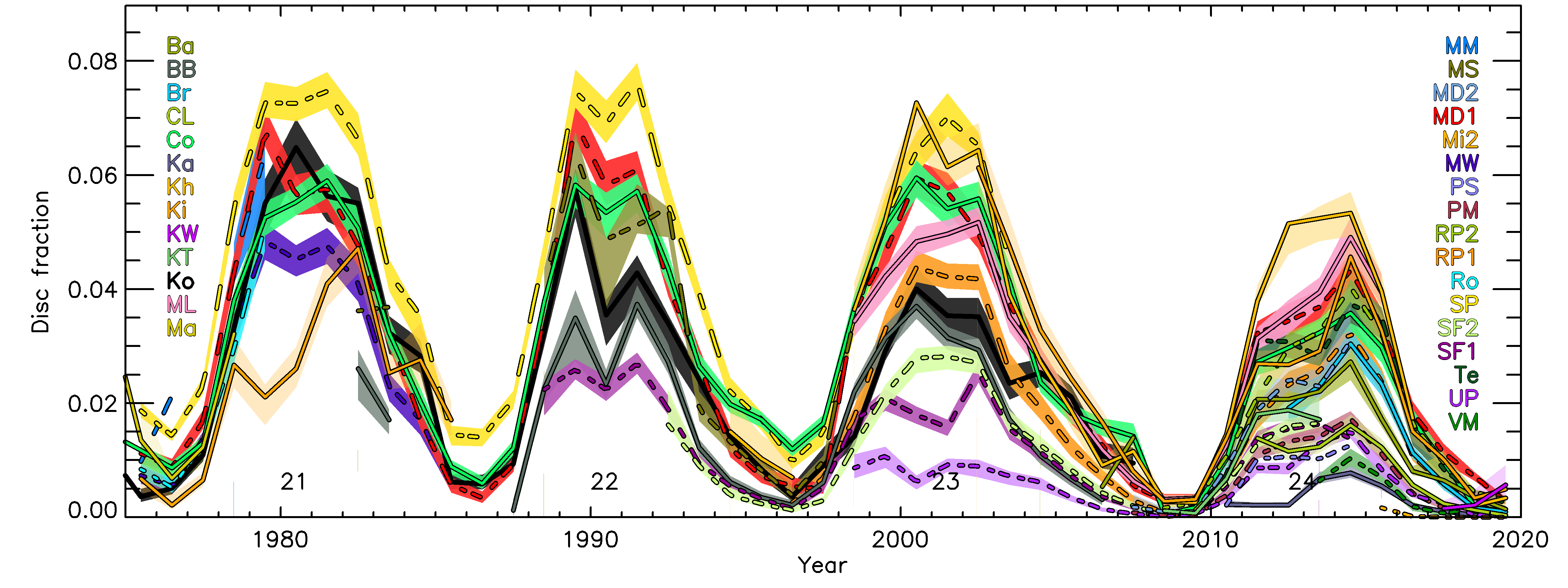}
	\caption{Evolution of plage areas given as fraction of the solar disc derived from analysis of the 38 datasets considered in this study. Shown are annual median values (lines) along with the asymmetric 1$\sigma$ interval (shaded surfaces) for each dataset as specified in the legend. To improve visibility, the archives listed in the legend in the right side of the plot are represented by dashed lines. Due to the scarcity of observations, the plage areas derived from the Ke/YR and CT observations in the top panel, as well as the Kh in the middle panel are represented by circles, triangles, and circles, respectively. The conventional solar cycle (SC) numbers are given below the curves.}
	\label{fig:dftimeplageothermoderndata}
\end{figure*}

\begin{figure*}
	\centering
	\begin{overpic}[width=1\textwidth]{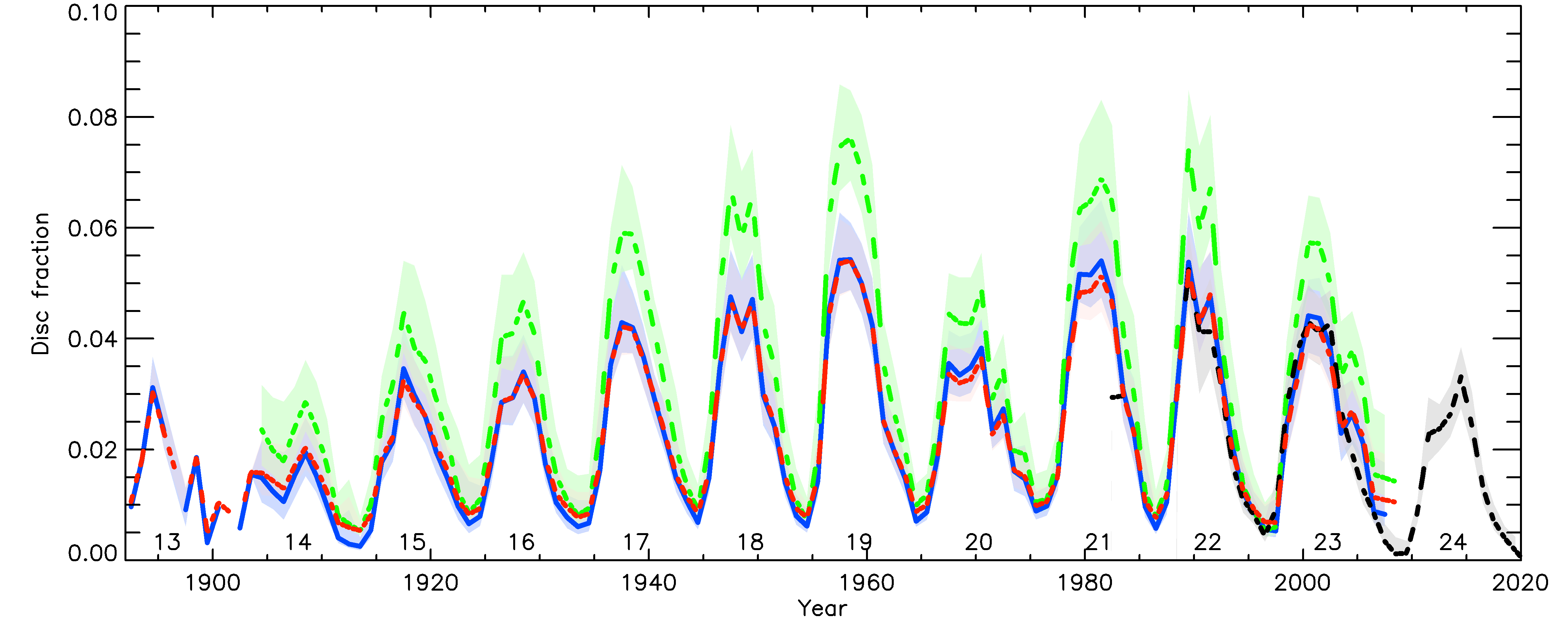}
		\put (10,31) {\tiny\textbf{RP1}}
		\put (10,35) {\tiny{\color{blue}\textbf{Ko}}}
		\put (10,33) {\tiny{\color{green}\textbf{MW}}}
		\put (10,37) {\tiny{\color{red}\textbf{Average of Ko and MW calibrated to RP1}}}
	\end{overpic}
	\caption{Backbone series of plage areas by using as reference the series from RP1 (dashed black), Ko (solid blue), MW (dashed green), and the average backbone of the MW and Ko series after their cross-calibration to the RP1 one (dashed red). Shown are annual median values (lines) along with the asymmetric 1$\sigma$ interval (shaded surfaces). The solar cycle numbers are given below the curves.}
	\label{fig:backbones}
\end{figure*}

\begin{figure*}
	\begin{overpic}[width=1\textwidth]{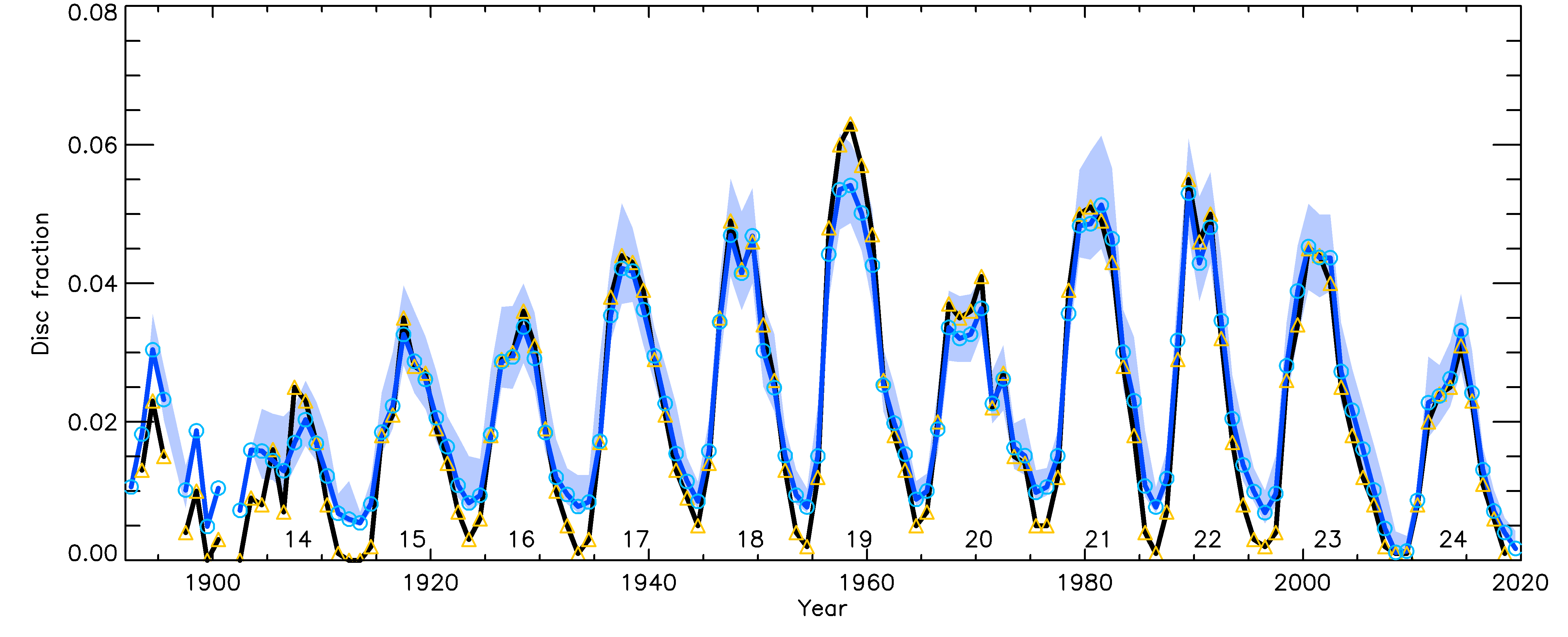}
		\put (10,35) {\tiny{\color{blue}\textbf{This study}}}
		\put (10,33) {\tiny\textbf{\cite{chatzistergos_analysis_2019}}}
	\end{overpic}
	\caption{Composite of plage areas in disc fraction derived in this study (blue line and ciel circles) along with the composite by \citet[][black line and yellow triangles]{chatzistergos_analysis_2019}. 
		Shown are annual median values (solid lines and symbols) along with the asymmetric 1$\sigma$ interval (shaded surfaces). The solar cycle numbers are given below the curves.}
	\label{fig:composite}
\end{figure*}

Tables \ref{tab:observatories} and \ref{tab:observatories2} illustrate the diversity of the analysed data in terms of, e.g., instruments, bandwidth used, central wavelength, and the resulting pixel scale of the images.
The observations from 14 of these datasets were stored on photographic plates (we will refer to all physical photographs as plates, even though celluloid film was used by some archives), a CCD camera was exclusively used for 25 datasets, while 4 datasets include data taken with a CCD camera as well as stored on photographic plates. 
We note that a few datasets include images obtained at the same observatory, but with different telescope or instrument, e.g. Kodaikanal and Kodaikanal Twin or Meudon spectroheliograms and Meudon filtergrams.
However, we do not make a distinction for the medium used to store the original data, whether it was a CCD camera or photographic plates. 
We note that this categorisation is merely to simplify the discussions in this manuscript. 
The arrangement of the datasets for the calibration procedure to produce the plage area composite is different and is outlined in Appendix \ref{sec:bakcboneasignemnt}.
The majority of the data stored on photographic plates were acquired with a spectroheliograph, with only Ro including images taken with an interference filter.
In contrast to that, most data taken with a CCD camera were obtained with interference filters, with only Co, MD1, Kh, and Ki images resulting from a spectroheliograph.
These four datasets include data taken with a CCD but also stored on photographic plates. 
We point also that MD1 data over 2002--2017 are provided in data cubes with observations taken at the core of the Ca~II~K line as well as centred at four different wavelengths away from the core.
In our derivation of the plage area series, we included only the observations taken at the core of the line (referred to as MD1) and the images taken at two extreme offsets (MDV and MDR, hereafter, for the data centred at the violet and red wing of the line). 
To discuss our results, we also analysed MD1 and Co observations centred beyond the K1 violet wing of the line (MDW and CoW, respectively) as well as ML data centred at the red wing of the line (MLW, hereafter).
All the off-band observations are summarised in Table \ref{tab:observatories2}.

The bandwidth of the analysed observations ranges between 0.1 and 9~\AA, thus sampling substantially different heights in the solar atmosphere.
These differences can be seen in Fig. \ref{fig:processedimages_raw}, where the images with relatively narrow bandwidth appear to have higher contrast in the plage regions, while the CLV is reduced compared to those with broader bandwidths.
In particular, quite indicative are the observations from MD1 and SF1 which were taken on the same day with nominal bandwidth of 0.15 and 9~\AA, respectively. 
Consequently, the network regions are enhanced in the MD1 image, while they are barely discernible in the one from SF1. 
Conversely, sunspots are clearly visible in the SF1 image, but barely hinted at the MD1 one.
Comparing RP1 and RP2, taken with a bandwidth of 2.5 and  1~\AA, respectively, the sunspots are only minutely reduced in size in RP2, while the CLV has also been reduced.

It is noteworthy that modern datasets include observations taken in general with broader bandwidths than the ones used for the historical data.
Figure \ref{fig:nominalbandwidths} shows the annual mean value of the nominal bandwidth from all datasets included in this study for which we have information on the bandwidth and for which the observations were centred at the core of the line. 
It is interesting that there is a slight decrease in the mean bandwidth between 1905 and 1920 while it remains roughly constant at around 0.3~\AA~up to the late 1980's. 
The variations over that period are rather low, with average bandwidths being in the range 0.2--0.5~\AA. However, the variations are more extreme since 1987, with a range of average bandwidths between 0.4 and 3.6~\AA.
To some degree this increase is due to the broad bandwidth of SF1 and SF2 data, but not entirely. Figure \ref{fig:nominalbandwidths} also shows the mean nominal bandwidth from all archives excluding SF1 and SF2. In this case there is still some, though weaker, increase of $\sim0.3$ \AA~in 1988 and an almost steady increase in the mean bandwidth after that.
Since the data availability during the mid 1980's is poorer than over other periods and the bandwidths change significantly, the uncertainty of cross-calibrating results from Ca~II~K data over that period is also higher.
However, we note that the central wavelength also affects the brightness of the magnetic features in the images and consequently their disc coverage.
We also note that some datasets appear to include observations taken slightly offset from the core of the Ca~II~K line.
This affects, for instance, the observations from ML, which have a central wavelength of 3934.15~\AA~instead of 3933.67~\AA.
Unfortunately, the precise values of the central wavelength used for the observations from most archives are not available, so that we cannot show its change with time.
Furthermore, the observations from Ka, PM, and VM have almost the same nominal bandwidth as RP1 (3.0, 2.5, 2.4, and 2.5~\AA, respectively), however, the CLV is stronger in the Ka, PM, and VM observations compared to the RP1 ones, hinting that Ka, PM, and VM observations might have been taken outside of the line centre or that the actual bandwidth is broader. 
A similar evaluation is more difficult for the historical data, which suffer from more artefacts than modern data. 
We only mention here that SP observations, taken with a nominal bandwidth of 0.5~\AA, have the lowest CLV among all datasets analysed in our work.
This suggests that the actual bandwidth used at SP might be narrower than the nominal value.

Furthermore, we note that in addition to a bandwidth different from the nominal one, there are other parameters of the observation that can affect the data.   
Indeed, parameters such as stray light contribution, blurring due to atmosphere seeing, over- or underexposure of photographic plates, vignetting, instrument- and setup-specific 
filter transmission profiles, potential contamination from secondary lobes in the filter transmission profile may have affected the CLV.
We point out that also modern data are affected by problems.
For instance, a few ML images taken with a CCD were found to be saturated.
Finally, observations from various sites have been copied and shared with other observatories. Therefore, it could also happen that images have been scanned and erroneously attributed to a wrong observatory.

It is also worth mentioning that some of the datasets listed in Table \ref{tab:observatories} underwent multiple digitisations, e.g. those from the Ko, Mi1, MW, and SP observatories, which were digitised with 8 and 16-bit devices.
Compared to the analysis by Paper 2, in this study we included 34 more datasets, as well as a new version of the Ko, MD1, and MW datasets. 
In particular, we used the Ko data from the more recent 16-bit digitisation, MD1 data over the period 1939--1948 and 1964--1967 which were not available before, and recently recovered data from the MW dataset.
The MW dataset stems from the 16-bit digitisation by \cite{lefebvre_solar_2005}, which is the same as the one used in Paper 2. 
The complete original dataset was, however,  considered lost due to a failure of the storage hardware. 
Luckily, it was recently recovered, and we found that it includes 3463 images which were missing from the dataset considered in Paper 2. 
However, 164 images from the dataset considered in Paper 2 are still missing in the new series. 
In this study, we analysed the recently recovered series of 16-bit MW data, but included the missing 164 images from the analysis by Paper 2.
Furthermore, observations from CT, Ko, Ma, Ro, and WS were found in 35mm celluloid films, which were distributed as the "Photographic journal of the Sun".
These were produced by the observatory of Rome over the period 1967--1978 as supplementary material to their monthly bulletins.
We digitised with 8-bit accuracy all the observations missing from our collection with the reflecta RPS 10M commercial film scanner, which is the same scanner previously used for the Ro observations \citep{chatzistergos_historical_2019}.
The datasets from CT, Ke, Kh, Ki, Ma, MD1, Sc, WS, and YR have only been partially digitised. 
Considering the large gaps in observations in the BB and MS datasets, it is possible that more data were taken over the 1980's, which, however, we were so far unable to recover. 

Finally, although the datasets from Ba, CL, Ka, PS, Te, and VM have multiple observations per day, for our analysis we used either the "best" observation of the day, as selected by the observers of the BK and VM datasets, or for the CL, Ka, PS, and Te datasets we used an automatic process to randomly select 1--3 images per day unaffected by cloud coverage. 
We did one exception for the first 7 days in June 2014.
For this period we analysed all available observations from all datasets to study the sensitivity of our results to daily variations of seeing (see Sec. \ref{sec:discussion} for more details).
The period was chosen randomly with the only requirements to be an active period and to be during summer for improved seeing conditions.
This way we provide a lower estimate of the uncertainty in the derived plage areas due to seeing variations.
Within these 7 days there are 36, 62, 89, 1640, 79, and 561 images in the Ba, Br, CL, Ka, ML, and Te datasets, respectively.

\subsection{Methods}

We consistently processed all images with the methods described by \cite{chatzistergos_analysis_2018,chatzistergos_analysis_2019,chatzistergos_historical_2020}. 
Briefly, we started by identifying the solar disc to extract the information on the coordinates of the disc centre and the radius \citep{chatzistergos_analysis_2019,chatzistergos_historical_2020}.
We applied the calibration for the digitisation device, where relevant data were available.
In particular, the Ko, KW, and MS datasets include information to perform the calibration of the CCD employed for the digitisation and observation, respectively. 
The CCD calibration has also been applied on BB (only over the period between 07 July 2000 and 21 September 2006), Co, KT, MD2, Mi2, RP1, and RP2 data, but it has not been applied, or it is unclear whether it has been applied, on the images from the other datasets.
However, the flat-field images taken at KW were found to exhibit large saturated areas, hence we decided not to use those flat-field files.
Here, we stress that the image calibration of the CCD recording device improves the accuracy of the analysed images and allows for removal of artefacts due to the device and its use, e.g. dust on the detector.
Figure \ref{fig:examplesflat} shows examples of the calibration of the CCD recording device for Ko, MS, and RP1 images. All the calibration data shown here exhibit intensity variations across the different quadrants of the CCD. Additionally, numerous dark small-scale round artefacts are evident in the MS observation. The flat image for the Ko observation also shows some scratch-like patterns. Such artefacts are accounted for by using the calibration images, thus reducing the uncertainties of analysing these data.
Besides, for all datasets we applied a data selection \citep{chatzistergos_analysis_2019} merely to exclude pathological cases with severely distorted discs, missing parts of the disc, or strong artefacts over the disc. 
Moreover, the solar disc in BB, Co, Kh, Ki, MD1, MW, SP, and Te images was re-sampled to account for its ellipticity following \citet{chatzistergos_historical_2020}.

Then, the images from the datasets with data stored on photographic plates were photometrically calibrated \citep{chatzistergos_analysis_2018} to account for the non-linear response of the photographic material.
All images, historical and modern, were compensated for the limb-darkening as described in Paper 1. 
In particular, we define a contrast image as $C_i=(I_i-I_i^{\mathrm{QS}})/I_i^{\mathrm{QS}}$, where $C_i$, $I_i$ are the contrast and intensity values of the calibrated image at pixel $i$, while $I_i^{\mathrm{QS}}$ is the intensity of the quiet Sun (QS, hereafter) at the pixel $i$.
Ky and YR data suffer from a very specific artefact manifesting itself as bright/dark arcs on the solar disc.
Therefore, to improve the accuracy of the analysis of the Ky and YR data we added one further processing step.
The process applied on these data is described in detail in Appendix \ref{sec:kyoto}.
Figure \ref{fig:processedimages_flat} displays examples of the calibrated and limb-darkening compensated contrast images after the preprocessing to correct for the elliptical discs (where needed) and to convert the historical images to density values for the same observations as shown in Fig. \ref{fig:processedimages_raw}. 
More details about the processing of the Ar, MM, MD1, Mi1, MW, RP1, Sc, and WS datasets can be found in Paper 2, about processing the CT and Ro datasets in  \cite{chatzistergos_historical_2019}, Ko in \cite{chatzistergos_delving_2019}, and Ky and SP in \cite{chatzistergos_historical_2020}.

All the processed images were segmented to identify plage areas with a multiplicative factor, $m_p=8.5$, to the standard deviation of the QS regions \citep{chatzistergos_analysis_2019}.
The segmentation was applied consistently with the same multiplicative factor to all the datasets.
The observation time for all archives was converted to Universal standard time (UT).
Figure \ref{fig:processedimages_mask} shows the corresponding segmentation masks of the observations shown in Fig. \ref{fig:processedimages_raw}, singling out the plage regions, which are the solar features mainly considered in this study.

\section{Analysis of the Ca~II~K series} 
\label{sec:results}
\subsection{Individual Ca~II~K series}

Figure \ref{fig:dftimeplageothermoderndata} shows the plage areas series derived from each dataset analysed in our study.
For clarity, we split the results into three panels, each one showing periods of roughly 4 solar cycles (SC).
The temporal profile of the evolution of plage areas is, in general, similar.
Thus, we can recognise the same features in all datasets during various periods, e.g. the period 1971--1975 or the double peaks of SC 22 maximum over the period 1989--1992.
However, there are also some obvious differences.

For instance, the plage areas from Ka, KW, Mi2, PM, UP, and VM are considerably lower than from all other datasets. 
This is in agreement with the comment in Sect. \ref{sec:data} that these data were probably taken off-band or with a broader bandwidth.
The plage areas from Co are considerably lower than from other datasets over SC 20. 
However, the annual values from Co data are not representative over that period due to the low number of Co images because of the relocation of the observatory \citep{lourenco_solar_2019}.
We also note that the Co plage areas over SC 19 are lower than SC 17 and 18, hinting for a potential issue with the data over SC 19.
Plage areas derived from Kh images are greater than those from Ko data over SC 23. 
This is contrary to the expectation considering that the bandwidth used at Kh is double the one used at Ko. 
Given that both observatories employed spectroheliographs\footnote{We remind that Kh used a CCD camera after 1994, while Ko used only photographic plates.}, this lends support to our suggestion that the actual bandwidth of the Ko observations is broader than reported or there is an offset in the central wavelength \citep{chatzistergos_analysis_2019,chatzistergos_delving_2019}.

We also compare the SF1 and SF2 series. 
These data have the same nominal observational characteristics except for the spatial resolution, which in SF1 data is half of that in SF2 data.
We find a linear correlation factor of 0.9 and RMS differences of 0.005 between the determined plage areas of the two series when considering only the 3821 days for which observations with both telescopes exist.
These differences are at least partly due to the lower resolution of SF1 data compared to the SF2 ones, which results in some smearing of the features.
However, this discrepancy might also be due to potential issues with SF1 data during 1997--2001.
This is evidenced by a sharp increase of plage areas over 1997 and a decrease during 2000--2001 around the activity maximum of SC 23.
When excluding the data between July 1997 and December 2001 we find a linear correlation factor of 0.95 and RMS differences of 0.003.

\begin{figure}
	\centering
	\includegraphics[width=1\linewidth]{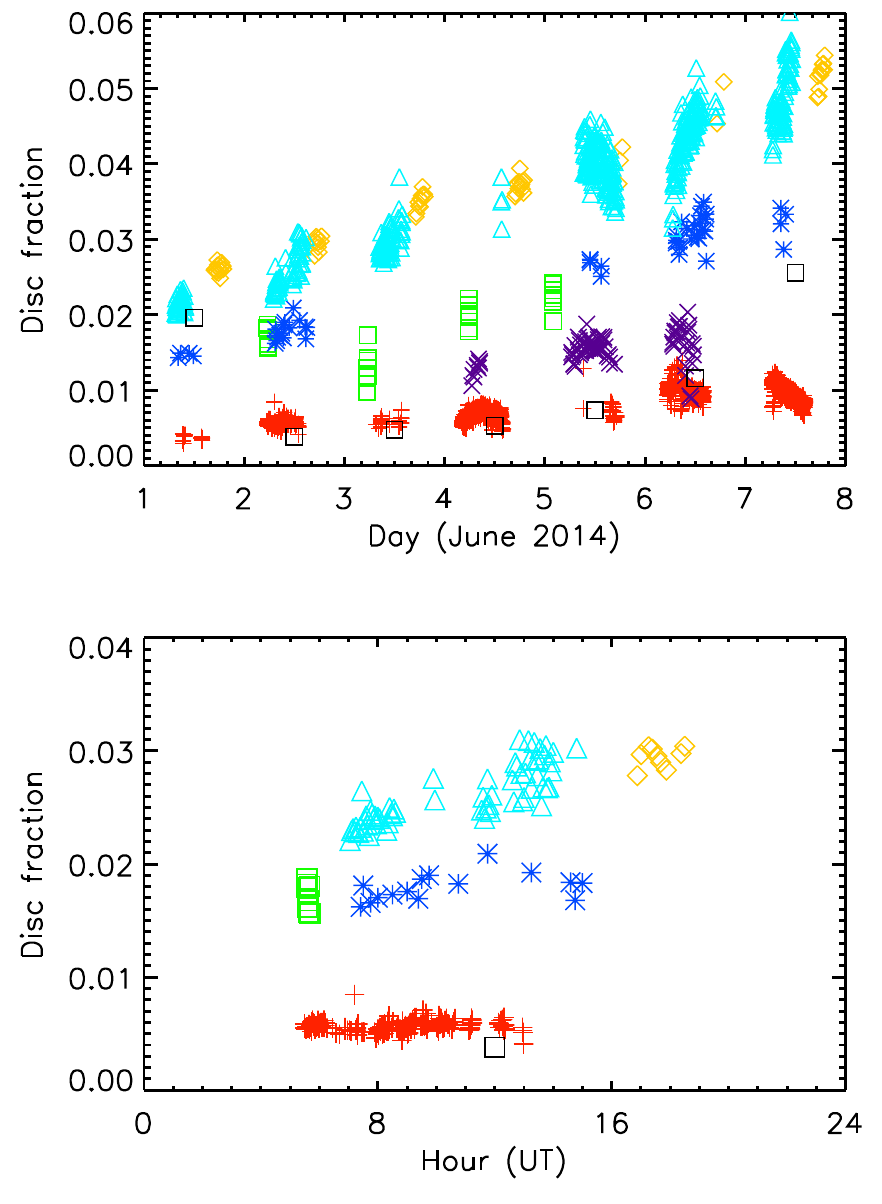}
	\caption{Plage areas in disc fraction as a function of time from the datasets of Ba (green squares), Br (blue asterisks), CL (purple x symbols, only in the top panel), Ka (red plus signs), ML (orange rhombuses), and Te (ciel triangles). Shown are results for individual images over the course of the first week in June 2014 (top panel) and over the course of 02 June 2014 (bottom panel). Also shown are the sunspot areas by \citet[][black squares]{balmaceda_homogeneous_2009} multiplied by 10 to bring them to roughly the same level as the plage areas for the sake of comparison. }
	\label{fig:june2014}
\end{figure}

\begin{figure}
	\centering
	\includegraphics[width=1\linewidth]{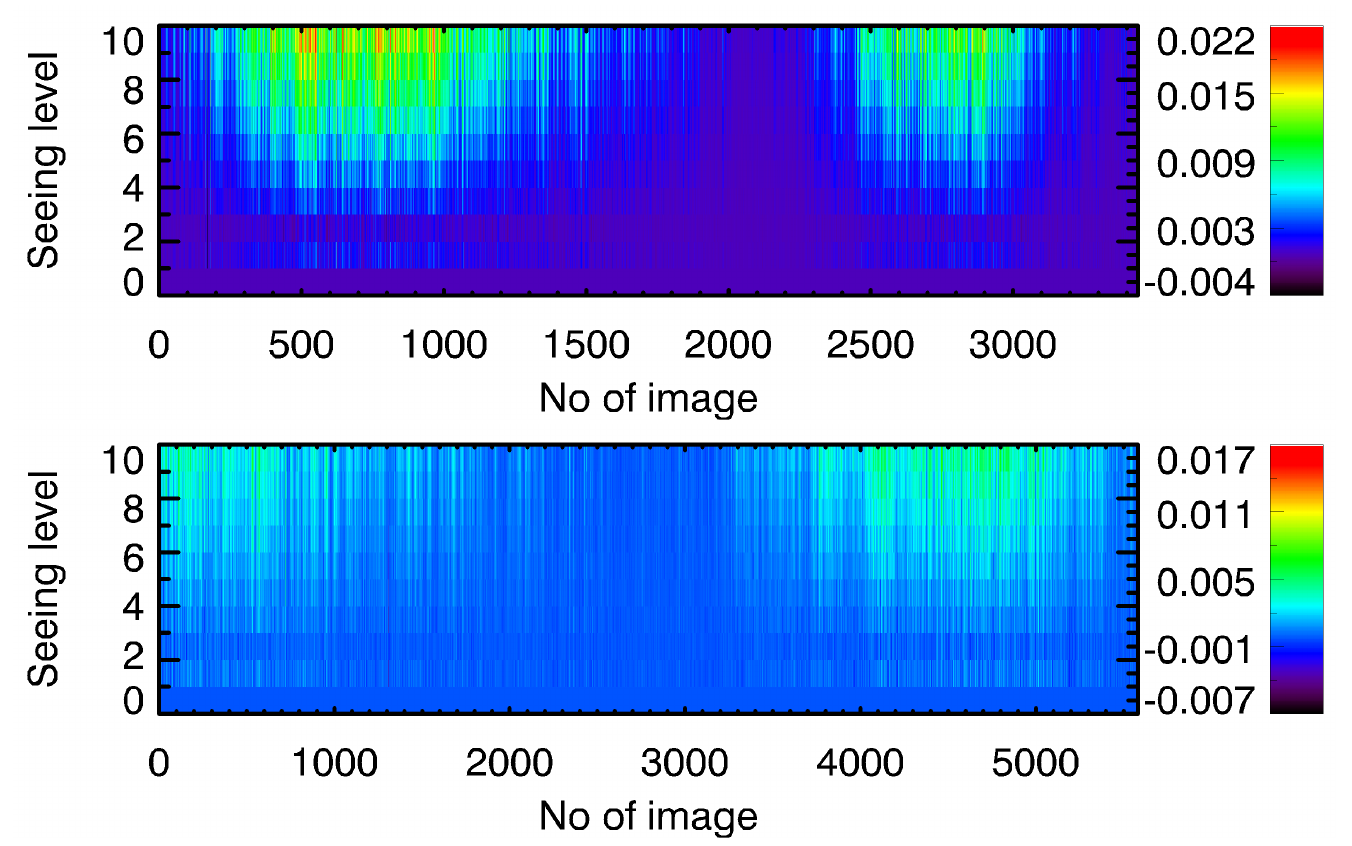}
	\caption{Colour-coded difference in fractional plage areas derived from RP1 (top) and MD1 (bottom) images downgraded to simulate effects of seeing and those derived from the original RP1 and MD1 images, respectively. The MD1 data used here are only those taken with a CCD camera between 2002 and 2017. Each row (column) of boxes shows results derived from a given observation (width of the smoothing Gaussian function). See Section \ref{sec:discussion} for details.}
	\label{fig:degradedpsptdiscfractionsseeingconstantkminimsigmaminithreshfaculae}
\end{figure}

\begin{figure*}
	\centering
	\includegraphics[width=1\linewidth]{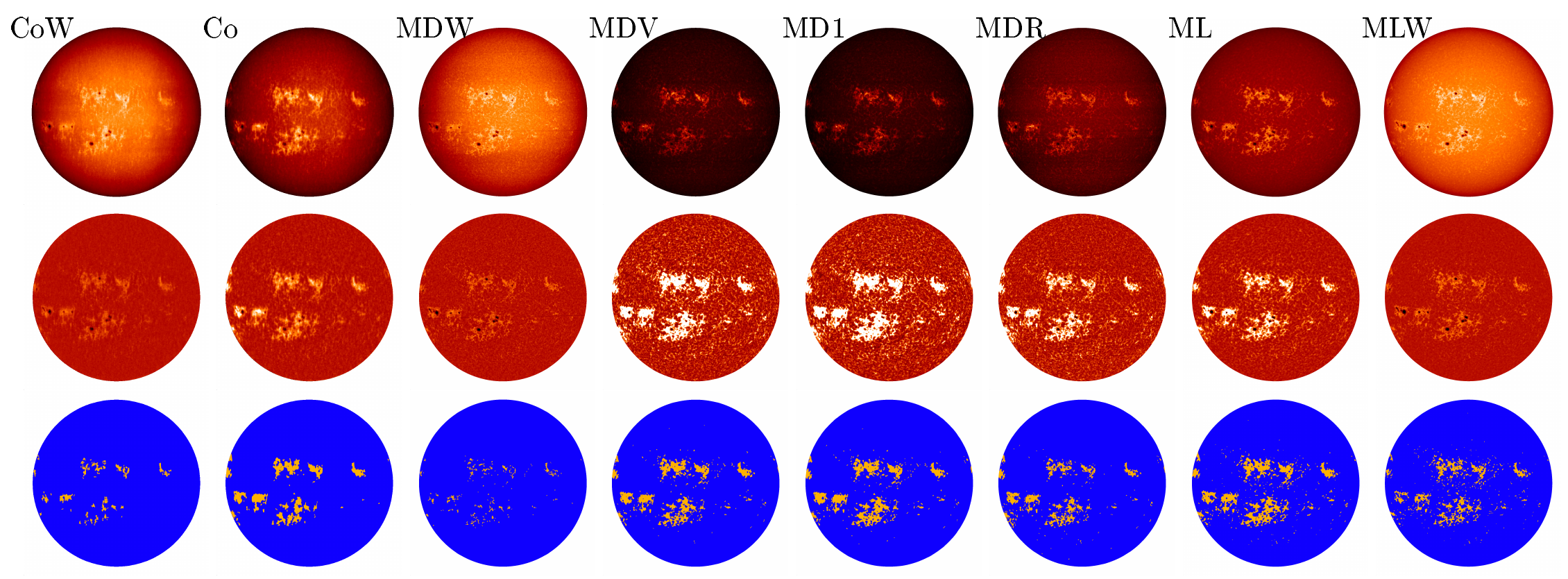}
	\caption{Examples of observations taken on the same date (04 July 2014) from the various datasets analysed in this study with images not centred at the core of the Ca~II~K line along with those from the same observatories centred at the core of the Ca~II~K line. Shown are images from the CoW, Co, MDW, MDV, MD1, MDR, ML, MLW datasets after the preprocessing to identify the disc and re-sample them to account for the disc ellipticity (top row), after CLV-compensation (middle row), and after the identification of plage (bottom row) . The images have been aligned to show the solar north pole at the top. The raw images are shown to their entire range of values, the CLV-compensated images are shown in the range [-0.5,0.5], while in the masks the plage regions are shown in orange and the quiet Sun and network regions in blue. We stress that the same threshold was used for all datasets to identify the plage regions, which is why the different datasets seem to give rather different plage coverage.}
	\label{fig:processedimages_offset}
\end{figure*}

\subsection{Plage areas composite}
\label{sec:composite}
In Paper 2 we presented a plage area composite derived from the analysis of 9 Ca~II~K archives.
The composite of plage areas was the average series of those obtained from using the results from the 8-bit Ko and 16-bit MW series as the references. 
In this study, we present a plage area composite obtained on the basis of the results from 38 datasets. 
These include the data from the 16-bit digitisation of the Ko dataset \citep{chatzistergos_delving_2019} as well as the recently recovered data from the 16-bit digitisation of MW. 
Furthermore, we use a different methodology to create our plage area composite series, employing the "backbone" approach \citep{svalgaard_reconstruction_2016,chatzistergos_new_2017}. 
In particular, we split the datasets into two backbones, roughly representing the historical and modern data separately. 
We took the RP1 series as the reference for the modern data backbone, while we use the Ko and MW series as references for the historical data. 
Appendix \ref{sec:bakcboneasignemnt} describes the assignment of the various analysed series to the backbones.

Following Paper 2 we cross-calibrated the individual series to the ones entering the backbone by using the daily statistics of the determined plage areas.
In particular, we started by computing daily mean plage areas for all series. 
Then, we constructed a probability distribution function (PDF, hereafter) matrix between each individual series and the corresponding backbone one. 
To create this matrix we first identified the days for which both series have a plage area measurement. 
Then we separated these days into arrays for which the secondary series reported plage areas in bins of 0.001 in disc fraction. 
For each of these arrays we computed the histogram of the reported plage areas from the backbone series in bins of 0.001 in disc fraction. 
For each of those arrays we normalised the histogram with the total number of data within that array, thus creating a PDF. 
See \cite{chatzistergos_new_2017} or paper 2 for more details on this process.
We computed the mean value of each PDF along with the asymmetric 1$\sigma$ interval, which we used to perform a weighted fit of a power law function with an offset \citep{chatzistergos_analysis_2019} and a linear fit. 
These relationships were used to scale the plage areas of the secondary series to the level of the backbone one.
Following Paper 2, we used the linear relation for the Ko and RP1 backbones, and  the power law for the MW ones.
All calibrated series were appended to the backbone one to create the backbone composite series. 
This way we construct one backbone for the modern data and two for the historical ones.
We cross-calibrated the two historical backbone series and the RP1 backbone series in the same way as done for the individual series. 
Then the two historical backbone series were averaged to create the average historical backbone series.

Figure \ref{fig:backbones} shows the RP1, Ko, MW, and the average historical backbone.
We note that the Ko and RP1 backbones are at similar levels over SCs 22 and 23, with the latter being slightly lower.
The variation of the plage areas in the MW backbone has higher amplitude than in the other backbone series by $\sim0.15$ in disc fraction.
The average historical backbone (average backbone series of Ko and MW after their cross-calibration to RP1) appears very similar to the raw Ko backbone, with most SCs after SC 20 being slightly reduced in amplitude.
We note that over SC 23 the plage areas in all backbone series except the RP1 exhibit a secondary peak around 2004. This is attributed to the Ko data over that period as was also mentioned by \cite{chatzistergos_delving_2019}. 

Figure \ref{fig:composite} shows the final composite\footnote{Available at \url{http://www2.mps.mpg.de/projects/sun-
		climate/data.html}} produced by merging the RP1 and the average historical backbones as well as the plage area composite obtained in Paper 2 for comparison. 
The two composites agree on the absolute level, which is expected since in this study we used RP1 as the reference, while the composite presented in Paper 2 had a scaling factor of 1 for RP1.
However, we notice that the plage areas at the maxima of SC 14, 19, 20, 22, and 23 are slightly reduced in the new composite with respect to those in Paper 2, while most minima are slightly elevated.
The increase of the values over activity minima is partly due to the inclusion of more data taken with a relatively narrow bandwidth compared to the composite in Paper 2.
Similarly, over SC 19, the newly added data favour lower activity level compared to that of MW or MD1 data.
However, we note that the majority of MD1 data over SC 19 are still not digitised, which can affect our composite series.

\begin{figure}
	\centering
	\includegraphics[width=1\linewidth]{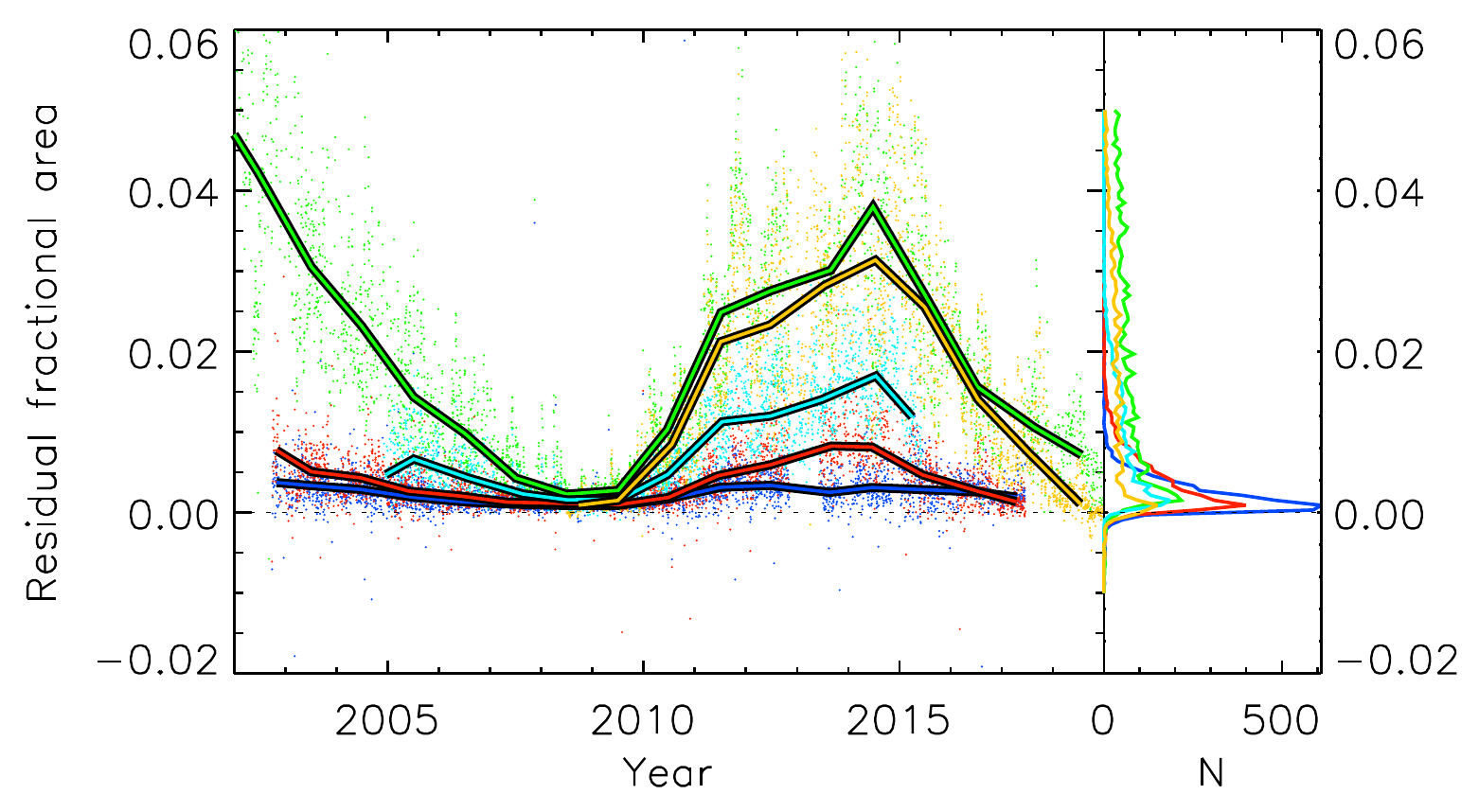}
	\caption{\textit{Left: }Difference of fractional plage area between those derived from centred and off-centred Ca~II~K observations as a function of time. The differences shown are for the Co-CoW (orange), MD1-MDW (green), MD1-MDV (blue), MD1-MDR (red), ML-MLW (ciel). Dots show daily values, while the solid lines show annual median values. The dotted horizontal line denotes a difference of 0. \textit{Right: } Distributions of the differences.}
	\label{fig:meudondifrentwavelengthcentre}
\end{figure}

\section{Discussion}
\label{sec:discussion}
Here we focus on the uncertainties of our results due to characteristics of the analysed data, i.e. seeing conditions, or central wavelength used for the observations, as well as those due to our method of producing  the composite plage area series.
The accuracy of the methods applied to the processing of the observations was tested and discussed by \cite{chatzistergos_analysis_2018,chatzistergos_analysis_2019,chatzistergos_delving_2019}.

First, we estimate the sensitivity of daily plage areas from individual datasets to the observational conditions.
These include the varying seeing conditions at different times of day and different locations, as well as stray-light.
To this end, we analysed all available observations of the first week of June 2014.
Figure \ref{fig:june2014} shows the derived plage areas over that period.  
Over those days, the plage areas from all datasets as well as the sunspot areas show a roughly steady increase. 
We also notice the derived plage areas to increase through the course of each day, which is consistent with the worsening seeing conditions towards local noon, that smears the plage regions.
However, at least part of this increase is due to the increase of solar activity over the course of that week.
This increase is expected to be greater for the archives with observations taken with narrow bandwidths, such as Te.
We find the areas from Ka to show more stable values within a given day, with only a slight increase of the areas around 06:00 UTC followed by a slight decrease after that.
The results in Fig. \ref{fig:june2014} allow a rough estimate of the uncertainty of the derived plage areas due to the seeing conditions during the acquisition of the images.
We find the areas from Ba, Br, and CL observations to exhibit a typical daily variation of $\sim0.01$  in disc fraction, with a somewhat lower variation for Te and Ka data (about 0.005), and a greater variation  (up to about 0.02) for ML data.
We cannot judge whether the bandwidth of the observations has any effect on these results.  

However, the passage of plage regions over the disc, part of which can go behind the limb or appearance of new plage regions at the limb within a day, also affects these results.
To remove this uncertainty, we also simulated the effects of seeing on RP1 and MD1 images by applying on them a Gaussian smoothing filter with varying widths. 
We used 10 values for the width uniformly distributed in the range [0,2] $\sigma$, where $\sigma$ is the standard deviation of the Gaussian function. 
Figure \ref{fig:degradedpsptdiscfractionsseeingconstantkminimsigmaminithreshfaculae} shows the residual plage fractional areas derived from the smoothed and the original (not smoothed) images.
Similarly to the results of the actual archives, we find a variation in the plage areas with generally higher values for the smoothed images.
This might come as a surprise considering that the plage areas are smoothed and hence their areas are expected to get reduced.
However, the smoothing also decreases the standard deviation of the QS regions, hence our threshold to isolate the plage areas is lower compared to the original images.
The variation in the residual plage areas follows the SC, with the highest values occurring during activity maxima.
The variations reach a value of 0.02 for the most severe case considered here.
The results for the MD1 and RP1 data are almost identical.
The only differences we identify are that MD1 data have marginally lower errors (up to 0.017 instead of 0.019 found in RP1 results when considering the common observations to MD1) and that there are images for which the plage areas decrease slightly more than RP1 ones (minimum value of -0.007 instead of -0.002 found in RP1 data for the common days to MD1).
The values for the differences in the plage areas are consistent with the results of the actual datasets.

We also estimate the uncertainties in the derived plage areas due to variations in the central wavelength of the observations.
All Ca II images can exhibit such variations, irrespectively of whether they were taken with a spectroheliograph, or with an interference filter.
In the case of the spectroheliograph, the variations are due to the positioning of the slit, e.g., because the observers intentionally or by mistake took observations off-centred from the line core.
In general, the cases when the observers intentionally took observations off-centred are documented, as in the archives from MD1 and Co for which observations centred at K1 and K3 are separated (the former are labelled MDW and CoW  respectively, in this paper, while the latter are abbreviated as MD1 and Co).
However, this practice was not always documented consistently for all of the datasets considered in our study.
In the case of the interference filters, the variations are more consistent and mostly due to the replacement of the filters or filter degradations, manifesting as a single offset or a drift in time, respectively.

\begin{figure*}
	\centering
	\begin{overpic}[width=1\linewidth]{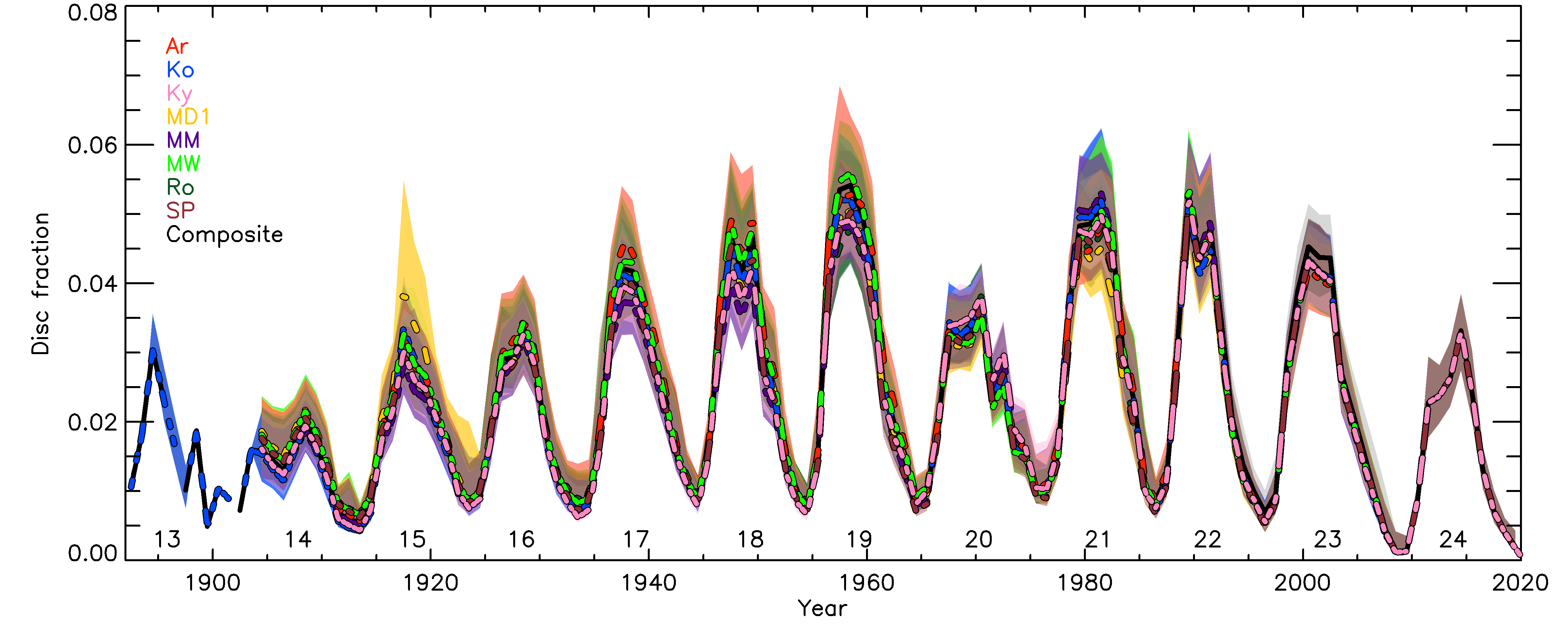}
		\put (15,35) {a)}
	\end{overpic}
	\begin{overpic}[width=1\linewidth]{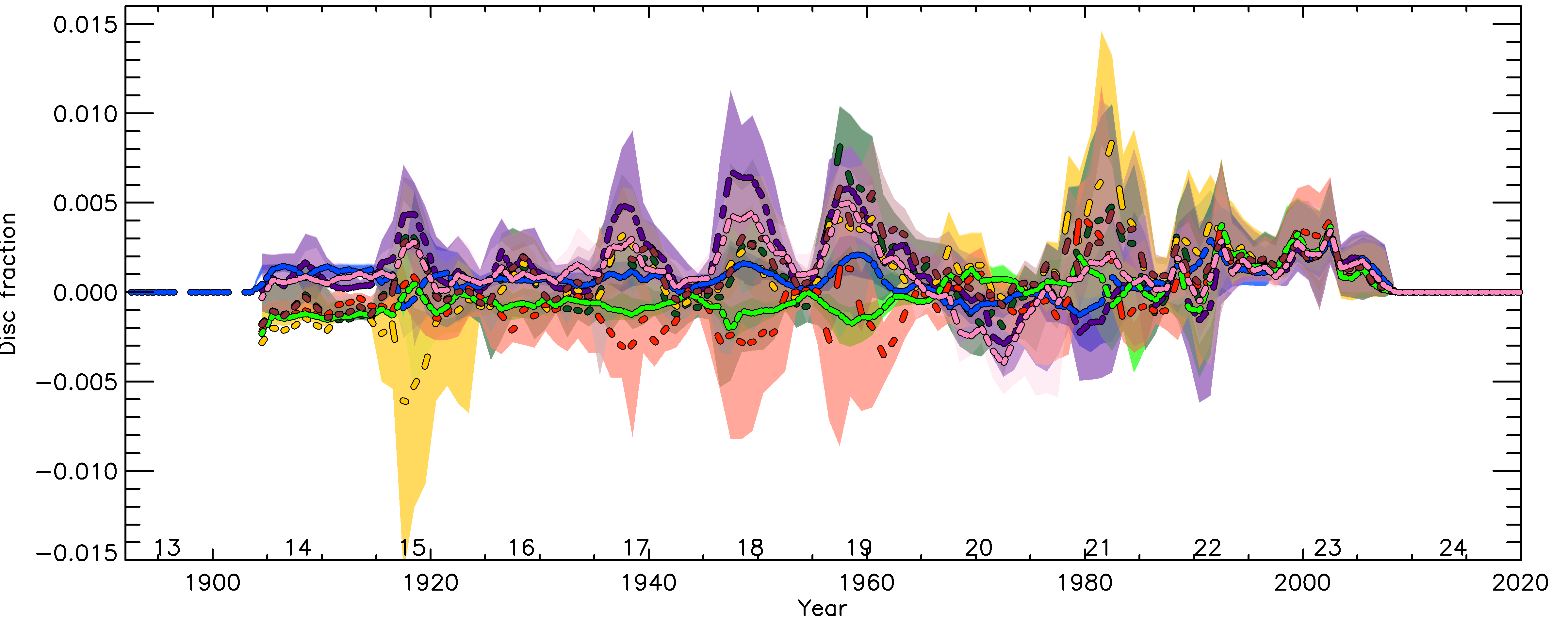}
		\put (15,35) {b)}
	\end{overpic}
	\caption{Plage area composites (a) and their difference to our main composite series (b) when the historical backbone series was computed with various individual historical datasets as the backbone reference instead of using the average of the calibrated Ko and MW ones to the RP1 one. RP1 is always taken as the overall reference for all composite series. The datasets used as backbones are those from  Ar (red), Ko (blue), Ky (pink), MD1 (orange), MM (purple), MW (green), Ro (dark green), and SP (brown) sites. The composite plage area series derived in this study is also displayed in black. Shown are the annual median values (lines) of the final plage area composites along with the asymmetric $1\sigma$ interval (shaded surfaces). The numbers below the curves denote the conventional SC numbering.}
	\label{fig:backbonesindividual2cl}
\end{figure*}

In order to test the effects of the different central wavelength on our derived plage areas we used the off-band observations from Coimbra, Meudon, and Mauna Loa.
Meudon has observations over the period 2002--2017 taken at four different wavelengths around the core of the Ca~II~K line along with the data taken at the core of the line.
All of those observations were taken with the same bandwidth.
Here we use only those taken at the centre of the line and the two extreme cases which correspond to offsets of $\pm$0.3~\AA.
We will refer to those series as MDV and MDR for the data taken in the violet and red part of the wing of the line, respectively.
Coimbra and Meudon have also observations centred beyond the K1V minimum, with an offset of -1.4~\AA~of the Ca~II~K line (CoW and MDW, hereafter), taken with the same bandwidth as those centred at the core of the line.
The Mauna Loa dataset includes data taken at the core of the Ca~II~K line as well as data taken 2.6~\AA~offset to the red wing (MLW, hereafter) of the line with a narrower bandwidth of 1~\AA~compared to those taken at the core of the line (which have a bandwidth of 2.7~\AA).

Figure \ref{fig:processedimages_offset} shows examples of these data for observations taken on 4 July 2014.
We notice that the contrast of plage areas decreases for all data taken off-centred, and unsurprisingly it is lower for data taken further away from the core of the line.
Moreover, the network regions are diminished in contrast and the sunspots are enlarged.
The MLW images are quite similar to the CoW and MDW ones, even though MLW is supposed to be taken further away from the core of the line than the other two.
However, the contrast of the MLW images is slightly greater than those from CoW and MDW.
This might be an effect of the differences in the bandwidth of these observations, with the one used for MLW being considerably broader than those used for CoW and MDW.

Figure \ref{fig:meudondifrentwavelengthcentre} gives the absolute difference in the derived plage areas between the datasets centred at the core of the line and those taken off-centred. 
The same segmentation method was applied to all the data.
The plage areas derived from such datasets also decreases compared to the values we get for the data centred at the core of the line.
The difference for the ML-MLW data is lower than for the MD1-MDW or Co-CoW ones, reaching a value of 0.02 during activity maximum.
The differences show variations following the SC, but a small offset is also noticed during activity minimum, being on average between 0.001 and 0.002 for MD1-MDV and MD1-MDW, respectively.
The differences are greatest for the cases MD1-MDW and Co-CoW reaching values of 0.04 during activity maxima.
We expect the typical variations of the central wavelength in the historical data to be similar to those for the cases of MD1-MDR and MD1-MDV, and hence they can provide a very approximate uncertainty level in the determined plage areas from the historical data due to shifts in the central wavelength.
For MD1-MDR and MD1-MDV we notice the distribution of the differences to be quite narrow, compared to the other cases tested here.
The differences in the derived plage areas for these cases are on average (RMS difference) less than 0.003 (0.003). 
However, we also notice that even though the wavelength offsets for MDR and MDV are equal in absolute value, the results for the plage areas are not exactly the same.
The difference in the plage areas for MDR are greater than for MDV.
This is unsurprising considering the similarity of MDV and MD1 images compared to the MDR ones (see Fig. \ref{fig:processedimages_offset}).

Next, we tested to which degree the composite series is affected by our choice of individual backbone datasets. 
In this process we used all historical datasets with sufficiently long periods to act as backbone references.
These are the Ar, Ko, Ky, MD1, MM, MW, Ro, and SP datasets.
Figure \ref{fig:backbonesindividual2cl} shows the resulting plage area composite series.
We did not consider Co, Mi1, or Mi2 in this test due to the poor overlap with many of the other datasets. 
Table \ref{tab:backbones} lists the assignment of the various plage series to the individual backbones.
For consistency, all series are normalized to the level of the RP1 backbone. 
All produced composites agree almost perfectly over SCs 22--24, owing to the RP1 backbone.
However, there are disagreements for the remaining cycles. 
The differences are greatest for the composites created with MM, Ky, and SP as the backbones for the historical data.
This is consistent with the rather low amount of data within those datasets, rendering the calibration of the various other (non-backbone) datasets to their level more uncertain.
The remaining reconstructions show results that are very similar to our proposed series.
The differences are typically below 0.005, but increase over SC 19 to 0.02 for daily values. 
This gives a rough estimate of the uncertainty in our official composite series due to the selection of individual datasets to act as the reference.
We note, however, that the overlap between most of the series used as backbones for this test is not optimum and is always worse compared to that of the Ko and MW series.
Furthermore, by averaging the Ko and MW backbone series we reduce the uncertainty due to the choice of the reference series.

\begin{figure}
	\centering
	\begin{overpic}[width=1\linewidth]{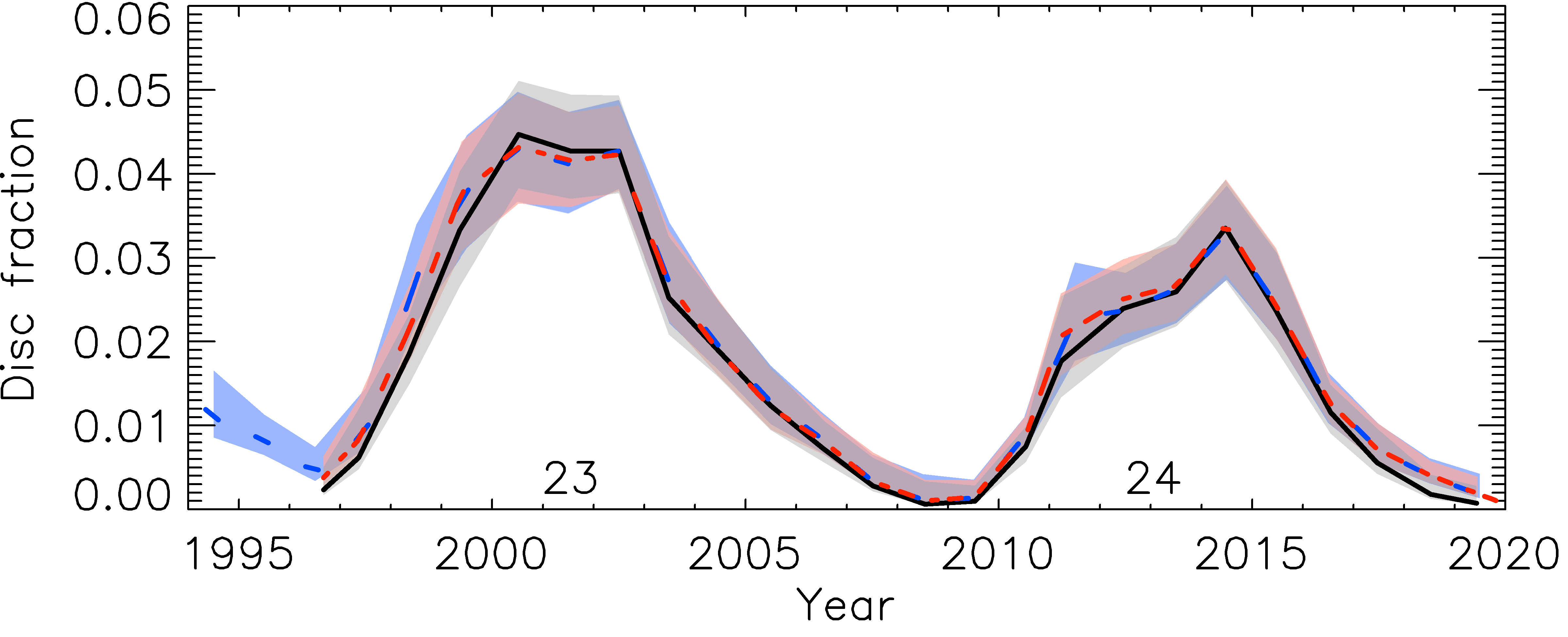}
		\put (15,35) {a)}
	\end{overpic}
		\begin{overpic}[width=1\linewidth]{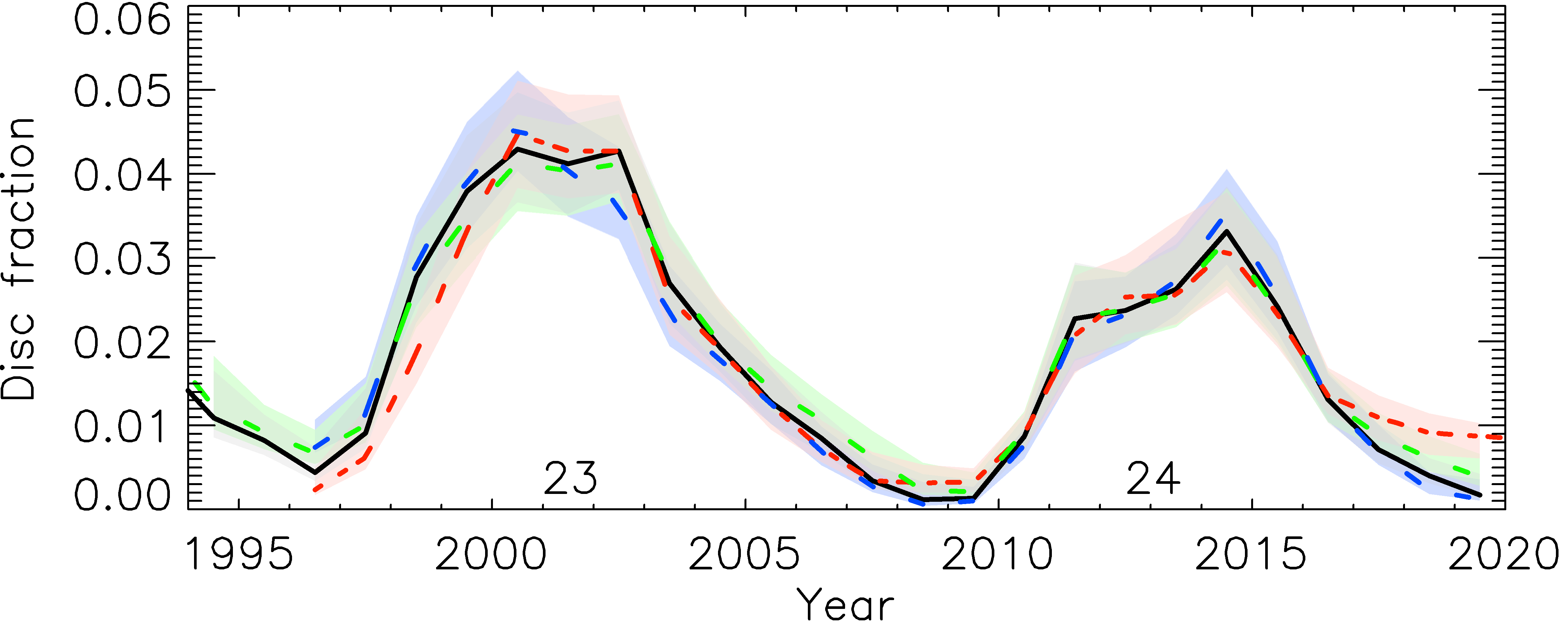}
		\put (15,35) {b)}
	\end{overpic}
	\caption{\textit{Top: }RP1 plage area series (solid black) plotted along with the RP1 backbone composite series used in our plage area composite (dashed blue) and RP1 backbone composite series keeping only the days common with the RP1 individual series (dashed red). \textit{Bottom: }RP1 backbone composite series when varying the included individual datasets based on their bandwidth. The different curves show the RP1 backbone composite by keeping Ba, Co, Kh, RP2, SF1, SF2, Te, Up (dashed green), Br, ML, BB (dashed blue) and Ka, PM, VM (dashed red). The RP1 backbone series used for our composite is shown in solid black. Depicted are annual median values (solid lines) along with the $1\sigma$ asymmetric intervals (shaded surfaces). The SC numbers are given below the curves.}
	\label{fig:romepsptbackbonesdb2cl}
\end{figure}

Figure \ref{fig:romepsptbackbonesdb2cl} a) shows the composite based on using  RP1 as a backbone series (blue). 
It is plotted along with the RP1 series on its own (black).
These two series match almost perfectly, with only small differences (RMS difference of 0.002 in disc fraction) which are greatest in 1998 (reaching 0.008 in disc fraction).
Also shown in Fig. \ref{fig:romepsptbackbonesdb2cl} a) is the composite using the RP1 backbone series when only the days included in the individual RP1 series are considered (red). 
In this case the differences are reduced, with an RMS difference of 0.001 and a maximum difference in 1999 of 0.003 in disc fraction.
This illustrates the accurate cross-calibration of the individual series to the RP1 backbone.

Finally, we also test the effect of including data taken with different bandwidths in the backbone series. 
For this purpose, we use the RP1 backbone and produce 3 different versions of it: i) keeping only datasets with either bandwidths narrower than 2~\AA or broader than 3~\AA~(Ba, CL, Co, Kh, PS, RP2, SF1, SF2, Te, Up). Since the RP1 bandwidth is 2.5~\AA, these limits imply that the chosen bandwidths differ by at least $\pm 0.5$~\AA~from that of RP1. ii) including only datasets with the nominal bandwidth between 2~\AA~and 3~\AA, i.e. within $\pm 0.5$~\AA~of the bandwidth of RP1. In this case, we further subdivide the datasets according to whether the bandwidths that have been assigned to them ii a) appear to be consistent with their actual behaviour (BB, Br, ML) ii b) or their assigned bandwidth does not appear to be consistent with their actual behaviour (Ka, PM, VM). 
For this test, we considered the BB and ML datasets only over the periods when the final instrumentation was used (see Appendix \ref{sec:bakcboneasignemnt}), in order to avoid inconsistencies due to instrumental changes. 
Figure \ref{fig:romepsptbackbonesdb2cl} b) displays the three test backbone series in comparison to the one used in our composite.
All three reconstructions of the RP1 backbone series lie within the uncertainties.
There are generally small differences, which are greater in 2011, before 2000, and after 2018, reaching up to 0.01 in disc fraction.

\section{Summary and conclusions}
\label{sec:conclusions}
We have processed 43 datasets of full-disc Ca~II~K  observations spanning the period 1892--2019 to derive the evolution of plage areas over the last 12 solar cycles.
We processed the data consistently with the method developed by \cite{chatzistergos_analysis_2018,chatzistergos_analysis_2019}.
An extra step in the processing was added to ensure accurate results from the analysis of images from the Ky and YR datasets, which suffer from specific artefacts along arcs on the solar disc. 
We adapted our processing such that these artefacts can be precisely accounted for and showed that we can obtain accurate results for those datasets as well.

We used our results for the 38 datasets with observations centred at the Ca~II~K line to create a plage area composite by applying the backbone approach employed by \cite{chatzistergos_new_2017} to create a sunspot group number composite.
We created two backbones, one mostly for the modern CCD-based data and another for the historical data, which were mainly stored on photographic plates. 
We considered the plage area series from Rome (RP1) observations as the overall reference series and to act as the backbone for the modern data, while both Kodaikanal (Ko) and Mt Wilson (MW) acted as the references for the historical data.
The obtained composite on the whole reasonably is consistent with the one we presented in Paper 2, although small differences exist.
The composite derived in this study has an average annual coverage of 98\% for the periods after 1906, with observations for only 672 days missing.
The coverage, however, remains rather low for the period 1892--1906 with 4917 days without any observations.
Previous results in the literature were based on considerably poorer temporal coverages.
We also illustrated the importance of using multiple datasets to improve the annual coverage in comparison to the case when the results derive from observations from the Ko and MW datasets alone, which are the ones most employed in studies of the plage areas evolution so far.

Many observatories,  whose data have been analysed here, have stopped the solar monitoring in the Ca~II~K line.
However, observations in the Ca~II~K line continue at the Ba, Br, CL, Co, Ka, Kh, Ki, KW, MD1, Mi2, PM, RP1, RP2, SF1, SF2, Te, UP, and VM sites to this day.
A combination of the data from all those sites provides a full annual coverage.
There is no day without an observation in our composite series since 2010.
However, there are still gaps in our composite.
Hopefully, more historical archives, such as those from Abastumani \citep{khetsuriani_solar_1981}, Anacapri \citep{antonucci_chromospheric_1977}, Cambridge \citep{moss_report_1942}, Crimea, Ebro \citep{curto_historical_2016}, Huairu \citep{suo_full-disk_2020}, Kandilli \citep{dizer_kandilli_1968}, Locarno \citep{waldmeier_swiss_1968}, Madrid \citep{vaquero_spectroheliographic_2007}, and the remaining data from the Baikal, Catania, Kenwood, Kharkiv, Kislovodsk, Manila, Meudon, Yerkes, Wendelstein, and Schauinsland sites will be digitised in the near future, which can potentially further increase the daily coverage of the data entering our composite series.

We have shone light to various issues affecting individual Ca~II~K archives.
We accounted for some of these in a simple manner by splitting the series into different parts and performing their cross-calibration to the backbone series separately.
However, there are more issues that remain unaccounted for in our analysis, e.g. the variable bandwidth and central wavelength of the observations taken with a spectroheliograph.
For these we have presented an estimate of the uncertainty in our results.
However, we plan to further address the effects due to variable bandwidth and central wavelength of the observations with a machine learning approach, considering that such methods have shown great potential on image-to-image conversion \citep[e.g.][]{kim_solar_2019,galvez_machine-learning_2019,park_generation_2019}.
Besides, we plan to continue processing the data from the currently operating programs in the Ca~II~K line to update the composite at regular intervals.
Finally, we plan to continue improving and updating the composite by including further historical data whenever these are digitized and made available to us.

\begin{acknowledgements}
The authors thank the observers at the Arcetri, Baikal, Big Bear, Brussels, Calern, Catania, Coimbra, Kanzelh\"ohe, Kharkiv, Kenwood, Kislovodsk, Kodaikanal, Kyoto, Manila, Mauna Loa, McMath-Hulbert, Mees, Meudon, Mitaka, Mt Wilson, Pic du Midi, Rome, Sacramento Peak, San Fernando, Schauinsland, Teide, Upice, Vala\v{s}sk\'{e} Mezi\v{r}i\v{c}\'{i}, Wendelstein, and Yerkes sites for all their work in carrying out the observing programs.
We thank Isabelle Buale for all her efforts to digitise the Meudon archive.
We thank Jeff Kuhn and Cindy Maui for locating and sharing with us the Mees Ca~II~K data.
We thank Satoru Ueno and Reizaburo Kitai for providing the Kyoto observations.
We thank the anonymous referee for the constructive comments that improved this manuscript.
We thank Ester Antonucci, Alexi Baker, Angie Cookson, Martina Exnerov\'a, Bernhard Fleck, Detlef Groote, Laurent Koechlin, Libor Len\v{z}a, Mustapha Meftah, Werner P\"otzi, Roger Ulrich, John Varsik, and Hubertus W\"ohl for providing information about various Ca~II~K data.
This work was supported by FP7 SOLID, and by the BK21 plus program through the National Research Foundation (NRF) funded by the Ministry of Education of Korea. This research has received funding from the European Union's Horizon 2020 research and innovation program under grant agreement No 824135 (SOLARNET). 
The Coimbra researchers thank the project  ReNATURE (CENTRO–01–0145–FEDER–000007–BPD16).
We acknowledge the "Observateurs associ\'es" for their commitment to image acquisition and processing; IRAP for the instrumental and database management; Universit\'e de Toulouse, CNRS, and Fiducial for the funding.
We acknowledge Paris Observatory for the use of spectroheliograms and the Royal Observatory of Belgium, Brussels for USET data. The Kanzelh\"ohe Ca~II~K data were provided by the Kanzelh\"ohe Observatory, University of Graz, Austria.
ChroTel is operated by the Kiepenheuer-Institute for Solar Physics in Freiburg, Germany, at the Spanish Observatorio del Teide, Tenerife, Canary Islands. The ChroTel filtergraph has been developed by the Kiepenheuer-Institute in cooperation with the High Altitude Observatory in Boulder, CO, USA.
We acknowledge \url{www.observethesun.com} and \url{www.solarstation.ru} for storing the Kislovodsk data.
This work used data provided by the MEDOC data and operations centre (CNES / CNRS / Univ. Paris-Sud).
The Kenwood observations used here are from lantern slides in the Division of History of Science and Technology at Yale University's Peabody Museum of Natural History (objects YPM HST.340744, HST.340745, HST.340747, and HST.340752).
The Yerkes observations are courtesy of the University of Chicago Photographic Archive, Special Collections Research Center, University of Chicago Library.
This research has made use of NASA's Astrophysics Data System.
\end{acknowledgements}

\bibliographystyle{aa}
\bibliography{_biblio1}   

\begin{thebibliography}{81}
\expandafter\ifx\csname natexlab\endcsname\relax\def\natexlab#1{#1}\fi

\bibitem[{Antonucci {et~al.}(1977)Antonucci, Azzarelli, Casalini, \&
  Cerri}]{antonucci_chromospheric_1977}
Antonucci, E., Azzarelli, L., Casalini, P., \& Cerri, S. 1977, Solar Physics,
  53, 519

\bibitem[{Balmaceda {et~al.}(2009)Balmaceda, Solanki, Krivova, \&
  Foster}]{balmaceda_homogeneous_2009}
Balmaceda, L.~A., Solanki, S.~K., Krivova, N.~A., \& Foster, S. 2009, Journal
  of Geophysical Research: Space Physics, 114

\bibitem[{Barata {et~al.}(2018)Barata, Carvalho, Dorotovič, Pinheiro, Garcia,
  Fernandes, \& Lourenço}]{barata_software_2018}
Barata, T., Carvalho, S., Dorotovič, I., {et~al.} 2018, Astronomy and
  Computing, 24, 70

\bibitem[{Belkina {et~al.}(1996)Belkina, Beletskij, Gretskij, \&
  Marchenko}]{belkina_ccd_1996}
Belkina, I.~L., Beletskij, S.~A., Gretskij, A.~M., \& Marchenko, G.~P. 1996,
  Kinematics and Physics of Celestial Bodies, 12, 55

\bibitem[{Berrilli {et~al.}(1999)Berrilli, Ermolli, Florio, \&
  Pietropaolo}]{berrilli_average_1999}
Berrilli, F., Ermolli, I., Florio, A., \& Pietropaolo, E. 1999, Astronomy and
  Astrophysics, 344, 965

\bibitem[{Bethge {et~al.}(2011)Bethge, Peter, Kentischer, Halbgewachs, Elmore,
  \& Beck}]{bethge_chromospheric_2011}
Bethge, C., Peter, H., Kentischer, T.~J., {et~al.} 2011, Astronomy and
  Astrophysics, 534, A105

\bibitem[{Chapman {et~al.}(1997)Chapman, Cookson, \&
  Dobias}]{chapman_solar_1997}
Chapman, G.~A., Cookson, A.~M., \& Dobias, J.~J. 1997, The Astrophysical
  Journal, 482, 541

\bibitem[{Chapman {et~al.}(2012)Chapman, Cookson, \&
  Preminger}]{chapman_comparison_2012}
Chapman, G.~A., Cookson, A.~M., \& Preminger, D.~G. 2012, Solar Physics, 276,
  35

\bibitem[{Chapman {et~al.}(2011)Chapman, Dobias, \&
  Arias}]{chapman_facular_2011}
Chapman, G.~A., Dobias, J.~J., \& Arias, T. 2011, The Astrophysical Journal,
  728, 150

\bibitem[{Chatterjee {et~al.}(2016)Chatterjee, Banerjee, \&
  Ravindra}]{chatterjee_butterfly_2016}
Chatterjee, S., Banerjee, D., \& Ravindra, B. 2016, The Astrophysical Journal,
  827, 87

\bibitem[{Chatterjee {et~al.}(2017)Chatterjee, Mandal, \&
  Banerjee}]{chatterjee_variation_2017}
Chatterjee, S., Mandal, S., \& Banerjee, D. 2017, The Astrophysical Journal,
  841, 70

\bibitem[{Chatzistergos(2017)}]{chatzistergos_analysis_2017}
Chatzistergos, T. 2017, Analysis of historical solar observations and long-term
  changes in solar irradiance, {PhD} thesis (Uni-edition)

\bibitem[{Chatzistergos {et~al.}(2019{\natexlab{a}})Chatzistergos, Ermolli,
  Falco, Giorgi, Guglielmino, Krivova, Romano, \&
  Solanki}]{chatzistergos_historical_2019}
Chatzistergos, T., Ermolli, I., Falco, M., {et~al.} 2019{\natexlab{a}}, in Il
  {Nuovo} {Cimento}, Vol. 42C, 5

\bibitem[{Chatzistergos {et~al.}(2018{\natexlab{a}})Chatzistergos, Ermolli,
  Krivova, \& Solanki}]{chatzistergos_ca_2018}
Chatzistergos, T., Ermolli, I., Krivova, N.~A., \& Solanki, S.~K.
  2018{\natexlab{a}}, in {IAU} {Symposium}, Vol. 340, Long-term {Datasets} for
  the {Understanding} of {Solar} and {Stellar} {Magnetic} {Cycles}, ed.
  D.~Banerjee, J.~Jiang, K.~Kusano, \& S.~Solanki (Cambridge, UK: Cambridge
  University Press), 125--128

\bibitem[{Chatzistergos {et~al.}(2019{\natexlab{b}})Chatzistergos, Ermolli,
  Krivova, \& Solanki}]{chatzistergos_analysis_2019}
Chatzistergos, T., Ermolli, I., Krivova, N.~A., \& Solanki, S.~K.
  2019{\natexlab{b}}, Astronomy \& Astrophysics, 625, A69

\bibitem[{Chatzistergos {et~al.}(2020)Chatzistergos, Ermolli, Krivova, \&
  Solanki}]{chatzistergos_historical_2020}
Chatzistergos, T., Ermolli, I., Krivova, N.~A., \& Solanki, S.~K. 2020, in
  {IOP} {Journal} of {Physics}: {Conference} {Series} (Rome: accepted)

\bibitem[{Chatzistergos {et~al.}(2016)Chatzistergos, Ermolli, Solanki, \&
  Krivova}]{chatzistergos_exploiting_2016}
Chatzistergos, T., Ermolli, I., Solanki, S.~K., \& Krivova, N.~A. 2016, in
  Astronomical {Society} of the {Pacific} {Conference} {Series}, Vol. 504,
  Coimbra {Solar} {Physics} {Meeting}: {Ground}-based {Solar} {Observations} in
  the {Space} {Instrumentation} {Era}, ed. I.~Dorotovic, C.~E. Fischer, \&
  M.~Temmer, San Francisco, 227--231

\bibitem[{Chatzistergos {et~al.}(2018{\natexlab{b}})Chatzistergos, Ermolli,
  Solanki, \& Krivova}]{chatzistergos_analysis_2018}
Chatzistergos, T., Ermolli, I., Solanki, S.~K., \& Krivova, N.~A.
  2018{\natexlab{b}}, Astronomy \& Astrophysics, 609, A92

\bibitem[{Chatzistergos {et~al.}(2019{\natexlab{c}})Chatzistergos, Ermolli,
  Solanki, Krivova, Banerjee, Jha, \& Chatterjee}]{chatzistergos_delving_2019}
Chatzistergos, T., Ermolli, I., Solanki, S.~K., {et~al.} 2019{\natexlab{c}},
  Solar Physics, 294, 145

\bibitem[{Chatzistergos {et~al.}(2019{\natexlab{d}})Chatzistergos, {Ermolli,
  Ilaria}, {Solanki, Sami K.}, {Krivova, Natalie A.}, {Giorgi, Fabrizio}, \&
  {Yeo, Kok Leng}}]{chatzistergos_recovering_2019}
Chatzistergos, T., {Ermolli, Ilaria}, {Solanki, Sami K.}, {et~al.}
  2019{\natexlab{d}}, Astronomy \& Astrophysics, 626, A114

\bibitem[{Chatzistergos {et~al.}(2017)Chatzistergos, Usoskin, Kovaltsov,
  Krivova, \& Solanki}]{chatzistergos_new_2017}
Chatzistergos, T., Usoskin, I.~G., Kovaltsov, G.~A., Krivova, N.~A., \&
  Solanki, S.~K. 2017, Astronomy \& Astrophysics, 602, A69

\bibitem[{Curto {et~al.}(2016)Curto, Solé, Genescà, Blanca, \&
  Vaquero}]{curto_historical_2016}
Curto, J.~J., Solé, J.~G., Genescà, M., Blanca, M.~J., \& Vaquero, J.~M.
  2016, Solar Physics

\bibitem[{Dizer(1968)}]{dizer_kandilli_1968}
Dizer, M. 1968, Solar Physics, 3, 491

\bibitem[{Dorotovič {et~al.}(2010)Dorotovič, Rybák, Garcia, \&
  Journoud}]{dorotovic_north-south_2010}
Dorotovič, I., Rybák, J., Garcia, A., \& Journoud, P. 2010, 20th National
  Solar Physics Meeting, 20, 58

\bibitem[{Ermolli {et~al.}(2003)Ermolli, Berrilli, \&
  Florio}]{ermolli_measure_2003}
Ermolli, I., Berrilli, F., \& Florio, A. 2003, Astronomy and Astrophysics, 412,
  857

\bibitem[{Ermolli {et~al.}(2018)Ermolli, Chatzistergos, Krivova, \&
  Solanki}]{ermolli_potential_2018}
Ermolli, I., Chatzistergos, T., Krivova, N.~A., \& Solanki, S.~K. 2018, in
  {IAU} {Symposium}, Vol. 340, Long-term {Datasets} for the {Understanding} of
  {Solar} and {Stellar} {Magnetic} {Cycles}, ed. D.~Banerjee, J.~Jiang,
  K.~Kusano, \& S.~Solanki (Cambridge, UK: Cambridge University Press),
  115--120

\bibitem[{Ermolli {et~al.}(2007)Ermolli, Criscuoli, Centrone, Giorgi, \&
  Penza}]{ermolli_photometric_2007}
Ermolli, I., Criscuoli, S., Centrone, M., Giorgi, F., \& Penza, V. 2007,
  Astronomy and Astrophysics, 465, 305

\bibitem[{Ermolli {et~al.}(2011)Ermolli, Criscuoli, \&
  Giorgi}]{ermolli_recent_2011}
Ermolli, I., Criscuoli, S., \& Giorgi, F. 2011, Contributions of the
  Astronomical Observatory Skalnate Pleso, 41, 73

\bibitem[{Ermolli {et~al.}(2010)Ermolli, Krivova, \&
  Solanki}]{ermolli_analysis_2010}
Ermolli, I., Krivova, N., \& Solanki, S.~K. 2010, in {COSPAR} {Meeting},
  Vol.~38, 38th {COSPAR} {Scientific} {Assembly}, 133

\bibitem[{Ermolli {et~al.}(2009{\natexlab{a}})Ermolli, Marchei, Centrone,
  Criscuoli, Giorgi, \& Perna}]{ermolli_digitized_2009}
Ermolli, I., Marchei, E., Centrone, M., {et~al.} 2009{\natexlab{a}}, Astronomy
  and Astrophysics, 499, 627

\bibitem[{Ermolli {et~al.}(2013)Ermolli, Matthes, Dudok~de Wit, Krivova,
  Tourpali, Weber, Unruh, Gray, Langematz, Pilewskie, Rozanov, Schmutz,
  Shapiro, Solanki, \& Woods}]{ermolli_recent_2013}
Ermolli, I., Matthes, K., Dudok~de Wit, T., {et~al.} 2013, Atmospheric
  Chemistry \& Physics, 13, 3945

\bibitem[{Ermolli {et~al.}(2009{\natexlab{b}})Ermolli, Solanki, Tlatov,
  Krivova, Ulrich, \& Singh}]{ermolli_comparison_2009}
Ermolli, I., Solanki, S.~K., Tlatov, A.~G., {et~al.} 2009{\natexlab{b}}, The
  Astrophysical Journal, 698, 1000

\bibitem[{Fontenla \& Landi(2018)}]{fontenla_bright_2018}
Fontenla, J.~M. \& Landi, E. 2018, The Astrophysical Journal, 861, 120

\bibitem[{Galvez {et~al.}(2019)Galvez, Fouhey, Jin, Szenicer, Muñoz-Jaramillo,
  Cheung, Wright, Bobra, Liu, Mason, \& Thomas}]{galvez_machine-learning_2019}
Galvez, R., Fouhey, D.~F., Jin, M., {et~al.} 2019, The Astrophysical Journal
  Supplement Series, 242, 7, publisher: American Astronomical Society

\bibitem[{Garcia {et~al.}(2011)Garcia, Sobotka, Klvana, \&
  Bumba}]{garcia_synoptic_2011}
Garcia, A., Sobotka, M., Klvana, M., \& Bumba, V. 2011, Contributions of the
  Astronomical Observatory Skalnate Pleso, 41, 69

\bibitem[{Golovko {et~al.}(2002)Golovko, Golubeva, Grechnev, Myachin, Trifonov,
  \& Khlystova}]{golovko_data_2002}
Golovko, A.~A., Golubeva, E.~M., Grechnev, V.~V., {et~al.} 2002, in Solar
  {Variability}: {From} {Core} to {Outer} {Frontiers}, Vol. 506 (ESA
  Publications Division), 929--932

\bibitem[{Gray {et~al.}(2010)Gray, Beer, Geller, Haigh, Lockwood, Matthes,
  Cubasch, Fleitmann, Harrison, Hood, Luterbacher, Meehl, Shindell, van Geel,
  \& White}]{gray_solar_2010}
Gray, L.~J., Beer, J., Geller, M., {et~al.} 2010, Reviews of Geophysics, 48,
  4001

\bibitem[{Haigh(2007)}]{haigh_sun_2007}
Haigh, J.~D. 2007, Living Reviews in Solar Physics, 4, 2

\bibitem[{Hale(1893)}]{hale_solar_1893}
Hale, G.~E. 1893, Memorie della Societa Degli Spettroscopisti Italiani, 21, 68

\bibitem[{Hale \& Ellerman(1903)}]{hale_rumford_1903}
Hale, G.~E. \& Ellerman, F. 1903, Publications of the Yerkes Observatory, 3,
  I.1

\bibitem[{Hanaoka(2013)}]{hanaoka_long-term_2013}
Hanaoka, Y. 2013, Journal of Physics Conference Series, 440, 2041

\bibitem[{Hanaoka \& {Solar Observatory of NAOJ}(2016)}]{hanaoka_past_2016}
Hanaoka, Y. \& {Solar Observatory of NAOJ}. 2016, in Astronomical {Society} of
  the {Pacific} {Conference} {Series}, Vol. 504, Coimbra {Solar} {Physics}
  {Meeting}: {Ground}-based {Solar} {Observations} in the {Space}
  {Instrumentation} {Era}, ed. I.~Dorotovic, C.~E. Fischer, \& M.~Temmer, San
  Francisco, 313

\bibitem[{Hiremath {et~al.}(2020)Hiremath, Rozelot, Sarp, Kilcik, D.~G., \&
  Gurumath}]{hiremath_nearly_2020}
Hiremath, K.~M., Rozelot, J.~P., Sarp, V., {et~al.} 2020, The Astrophysical
  Journal, 891, 151

\bibitem[{Hirtenfellner-Polanec {et~al.}(2011)Hirtenfellner-Polanec, Temmer,
  Pötzi, Freislich, Veronig, \&
  Hanslmeier}]{hirtenfellner-polanec_implementation_2011}
Hirtenfellner-Polanec, W., Temmer, M., Pötzi, W., {et~al.} 2011, Central
  European Astrophysical Bulletin, 35, 205

\bibitem[{Khetsuriani(1981)}]{khetsuriani_solar_1981}
Khetsuriani, T.~S. 1981, Solar Physics, 69, 405

\bibitem[{Kim {et~al.}(2019)Kim, Park, Lee, Moon, Bae, Lim, Jang, Kim, Cho,
  Choi, \& Cho}]{kim_solar_2019}
Kim, T., Park, E., Lee, H., {et~al.} 2019, Nature Astronomy, 1

\bibitem[{Kitai {et~al.}(2013)Kitai, Ueno, Maehara, Shirakawa, Katoda, Hada,
  Tomita, Hayashi, Asai, Isobe, Goto, \& Yamashita}]{kitai_digital_2013}
Kitai, R., Ueno, S., Maehara, H., {et~al.} 2013, Data Science Journal, 12,
  WDS213

\bibitem[{Klimeš {et~al.}(1999)Klimeš, Bělik, Klimeš, \&
  Marková}]{klimes_simultaneous_1999}
Klimeš, J., J., Bělik, M., Klimeš, J., S., \& Marková, E. 1999, in {ESA}
  {Special} {Publication}, Vol. 446, 8th {SOHO} {Workshop}: {Plasma} {Dynamics}
  and {Diagnostics} in the {Solar} {Transition} {Region} and {Corona}, ed.
  J.~C. Vial \& B.~Kaldeich-Schü, 375

\bibitem[{Koechlin {et~al.}(2019)Koechlin, Dettwiller, Audejean, Valais, \&
  Ariste}]{koechlin_solar_2019}
Koechlin, L., Dettwiller, L., Audejean, M., Valais, M., \& Ariste, A.~L. 2019,
  Astronomy \& Astrophysics

\bibitem[{Kopp {et~al.}(2016)Kopp, Krivova, Wu, \& Lean}]{kopp_impact_2016}
Kopp, G., Krivova, N., Wu, C.~J., \& Lean, J. 2016, Solar Physics, 291, 2951

\bibitem[{Lefebvre {et~al.}(2005)Lefebvre, Ulrich, Webster, Varadi, Javaraiah,
  Bertello, Werden, Boyden, \& Gilman}]{lefebvre_solar_2005}
Lefebvre, S., Ulrich, R.~K., Webster, L.~S., {et~al.} 2005, Memorie della
  Societa Astronomica Italiana, 76, 862

\bibitem[{Lenza {et~al.}(2014)Lenza, Srba, Gregorova, Exnerova, \&
  Lenzova}]{lenza_system_2014}
Lenza, L., Srba, J., Gregorova, B., Exnerova, M., \& Lenzova, N. 2014, System
  for simultaneous observation of solar flares in spectral lines of {H}-alpha
  and {CaII} {K}

\bibitem[{Loukitcheva {et~al.}(2009)Loukitcheva, Solanki, \&
  White}]{loukitcheva_relationship_2009}
Loukitcheva, M., Solanki, S.~K., \& White, S.~M. 2009, Astronomy and
  Astrophysics, 497, 273

\bibitem[{Lourenço {et~al.}(2019)Lourenço, Carvalho, Barata, Garcia,
  Carrasco, \& Peixinho}]{lourenco_solar_2019}
Lourenço, A., Carvalho, S., Barata, T., {et~al.} 2019, Open Astronomy, 28, 165

\bibitem[{Malherbe \& Dalmasse(2019)}]{malherbe_new_2019}
Malherbe, J.-M. \& Dalmasse, K. 2019, Solar Physics, 294, 52

\bibitem[{Meftah {et~al.}(2018)Meftah, Corbard, Hauchecorne, Morand, Ikhlef,
  Chauvineau, Renaud, Sarkissian, \& Damé}]{meftah_solar_2018}
Meftah, M., Corbard, T., Hauchecorne, A., {et~al.} 2018, Astronomy \&
  Astrophysics, 616, A64

\bibitem[{Meftah {et~al.}(2014)Meftah, Hochedez, Irbah, Hauchecorne, Boumier,
  Corbard, Turck-Chièze, Abbaki, Assus, Bertran, Bourget, Buisson, Chaigneau,
  Damé, Djafer, Dufour, Etcheto, Ferrero, Hersé, Marcovici, Meissonnier,
  Morand, Poiet, Prado, Renaud, Rouanet, Rouzé, Salabert, \&
  Vieau}]{meftah_picard_2014}
Meftah, M., Hochedez, J.-F., Irbah, A., {et~al.} 2014, Solar Physics, 289, 1043

\bibitem[{Miller(1965)}]{miller_new_1965}
Miller, R.~A. 1965, Applied Optics, 4, 1085

\bibitem[{Mohler \& Dodson(1968)}]{mohler_mcmath-hulbert_1968}
Mohler, O.~C. \& Dodson, H.~W. 1968, Solar Physics, 5, 417

\bibitem[{Moss(1942)}]{moss_report_1942}
Moss, W. 1942, Monthly Notices of the Royal Astronomical Society, 102, 86

\bibitem[{Naqvi {et~al.}(2010)Naqvi, Marquette, Tritschler, \&
  Denker}]{naqvi_big_2010}
Naqvi, M.~F., Marquette, W.~H., Tritschler, A., \& Denker, C. 2010,
  Astronomische Nachrichten, 331, 696

\bibitem[{Park {et~al.}(2019)Park, Moon, Lee, Kim, Lee, Lim, Shin, \&
  Kim}]{park_generation_2019}
Park, E., Moon, Y.-J., Lee, J.-Y., {et~al.} 2019, The Astrophysical Journal,
  884, L23

\bibitem[{Priyal {et~al.}(2014)Priyal, Singh, Ravindra, Priya, \&
  Amareswari}]{priyal_long_2014}
Priyal, M., Singh, J., Ravindra, B., Priya, T.~G., \& Amareswari, K. 2014,
  Solar Physics, 289, 137

\bibitem[{Pruthvi \& Ramesh(2015)}]{pruthvi_two-channel_2015}
Pruthvi, H. \& Ramesh, K.~B. 2015, International Conference on Optics and
  Photonics 2015, 9654, 96540I

\bibitem[{Raju(2018)}]{raju_temporal_2018}
Raju, K.~P. 2018, Monthly Notices of the Royal Astronomical Society, 478, 5056

\bibitem[{Rast {et~al.}(2008)Rast, Ortiz, \& Meisner}]{rast_latitudinal_2008}
Rast, M.~P., Ortiz, A., \& Meisner, R.~W. 2008, The Astrophysical Journal, 673,
  1209

\bibitem[{Shapiro {et~al.}(2017)Shapiro, Solanki, Krivova, Cameron, Yeo, \&
  Schmutz}]{shapiro_nature_2017}
Shapiro, A.~I., Solanki, S.~K., Krivova, N.~A., {et~al.} 2017, Nature
  Astronomy, 1, 612

\bibitem[{Singh \& Ravindra(2012)}]{singh_twin_2012}
Singh, J. \& Ravindra, B. 2012, Bulletin of the Astronomical Society of India,
  40

\bibitem[{Solanki {et~al.}(2013)Solanki, Krivova, \&
  Haigh}]{solanki_solar_2013-1}
Solanki, S.~K., Krivova, N.~A., \& Haigh, J.~D. 2013, Annual Review of
  Astronomy and Astrophysics, 51, 311

\bibitem[{Suo(2020)}]{suo_full-disk_2020}
Suo, L. 2020, Advances in Space Research, 65, 1054

\bibitem[{Svalgaard \& Schatten(2016)}]{svalgaard_reconstruction_2016}
Svalgaard, L. \& Schatten, K.~H. 2016, Solar Physics, 291, 2653

\bibitem[{Tlatov {et~al.}(2015)Tlatov, Dormidontov, Kirpichev, Pashchenko, \&
  Shramko}]{tlatov_synoptic_2015}
Tlatov, A.~G., Dormidontov, D.~V., Kirpichev, R.~V., Pashchenko, M.~P., \&
  Shramko, A.~D. 2015, Geomagnetism and Aeronomy, 55, 961

\bibitem[{Tlatov {et~al.}(2009)Tlatov, Pevtsov, \& Singh}]{tlatov_new_2009}
Tlatov, A.~G., Pevtsov, A.~A., \& Singh, J. 2009, Solar Physics, 255, 239

\bibitem[{Tlatov \& Tlatova(2019)}]{tlatov_polar_2019}
Tlatov, A.~G. \& Tlatova, K.~A. 2019, Geomagnetism and Aeronomy, 59, 6

\bibitem[{Vaquero {et~al.}(2007)Vaquero, Gallego, Acero, \&
  García}]{vaquero_spectroheliographic_2007}
Vaquero, J.~M., Gallego, M.~C., Acero, F.~J., \& García, J.~A. 2007, in
  Astronomical {Society} of the {Pacific} {Conference} {Series}, Vol. 368, The
  {Physics} of {Chromospheric} {Plasmas}, ed. P.~Heinzel, I.~Dorotovic, \&
  R.~J. Rutten

\bibitem[{Vogler {et~al.}(2008)Vogler, Brandt, Otruba, Pötzi, \&
  Hanslmeier}]{vogler_defects_2008}
Vogler, F., Brandt, P., Otruba, W., Pötzi, W., \& Hanslmeier, A. 2008, Central
  European Astrophysical Bulletin, 32, 141

\bibitem[{Waldmeier(1968)}]{waldmeier_swiss_1968}
Waldmeier, M. 1968, Solar Physics, 5, 423

\bibitem[{Wöhl(2005)}]{wohl_old_2005}
Wöhl, H. 2005, Hvar Observatory Bulletin, 29, 319

\bibitem[{Wu {et~al.}(2018)Wu, Krivova, Solanki, \& Usoskin}]{wu_solar_2018-2}
Wu, C.-J., Krivova, N.~A., Solanki, S.~K., \& Usoskin, I.~G. 2018, Astronomy \&
  Astrophysics, 620, A120

\bibitem[{Yeo {et~al.}(2017)Yeo, Krivova, \& Solanki}]{yeo_empire:_2017}
Yeo, K.~L., Krivova, N.~A., \& Solanki, S.~K. 2017, Journal of Geophysical
  Research: Space Physics, 2016JA023733

\bibitem[{Zuccarello {et~al.}(2011)Zuccarello, Contarino, \&
  Romano}]{zuccarello_solar_2011}
Zuccarello, F., Contarino, L., \& Romano, P. 2011, Contributions of the
  Astronomical Observatory Skalnate Pleso, 41, 85

\end{thebibliography}
\appendix

\section{Processing of Kyoto and Yerkes data}
\label{sec:kyoto}
\begin{figure*}
	\centering
\includegraphics[width=0.97\textwidth]{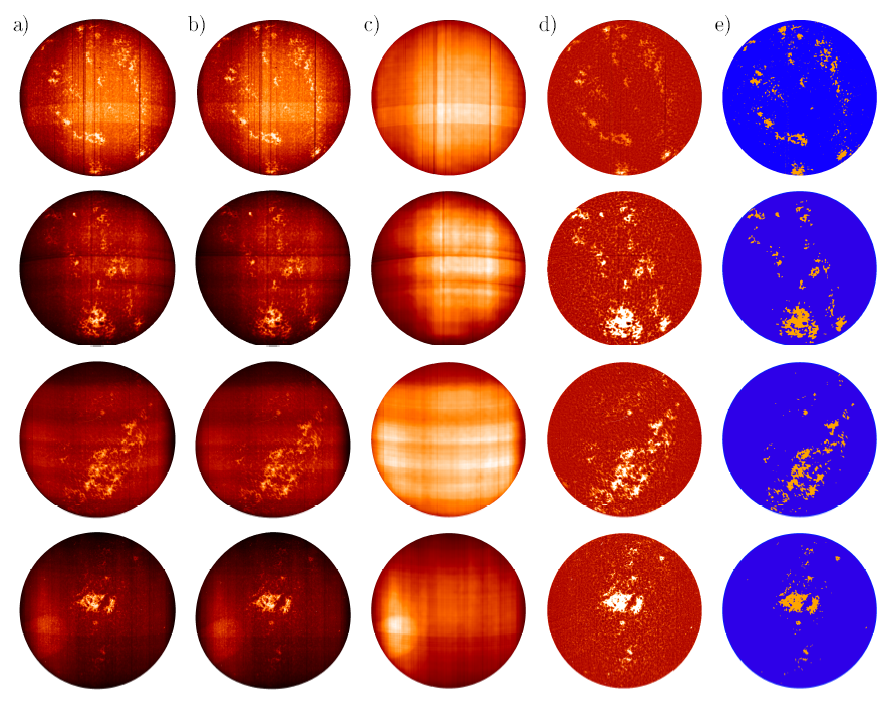}
\caption{Examples of processing steps applied on Ky observations taken on 22/10/1937 (1st row), 28/10/1939 (2nd row), 21/01/1959 (3rd row), and 30/04/1961 (4th row). Columns are: original raw density image (a), density image resampled to straighten the arcs (b), computed background of the original image (c), photometrically calibrated and limb-darkening-compensated image (d), and segmentation mask (e). The raw density images and the backgrounds are shown to the entire range of values within the disc, the calibrated images are shown in the range [-0.5,0.5], while the masks show plage regions in orange and QS and network regions in blue. The images are not compensated for ephemeris.} 
\label{fig:kyotoexampleprocessing}
\end{figure*}

Images from all sites that used spectroheliographs suffer from artefacts introduced by the motion of the slit of the spectroheliograph.
Most of these follow the slit's shape, which is linear in almost all instruments. 
For the Ky and YR archives, however, these artefacts appear to be curved rather than linear. 
To make things more complicated, there are linear artefacts roughly perpendicular to the curved ones as well (see Fig. 
\ref{fig:kyotoexampleprocessing}), likely introduced by other instrumental issues. 
The curved artefacts are more evident in the Ky data than in YR ones and for that reason we focus on Ky data here, even though the same process was applied on both series.
To account for the curved artefacts in the data, we adopted the following processing. 
During the pre-processing of the data, we identified the centre and the radius of the disc as well as the orientation and the curvature of the arcs within each raw image.
The identification of the linear and curved segments was initially done automatically. 
The linear segments in most images are aligned with the frame of the photograph. 
The frame, however, is not always aligned in the digital image.
Therefore, we applied Sobel filtering to identify the frame and then determine the angle needed to align the linear segments vertically.
The curved segments were identified by singling out bright or dark regions in the image after it was divided by a map constructed with a running window median filter with width of 10 pixels.

However, both the linear and the curved segments are not always clearly visible and the code was not always able to detect them. 
For that reason, the results were afterwards manually inspected and corrected when deemed necessary.
We identify pixels belonging to the same arc and for each pixel $n$ we determine its horizontal distance from the left side of the image, $y_n$, and its vertical distance from the bottom of the image, $x_n$.
In order to get the parameters of the arcs, we fit a parabolic function of the form:
\begin{equation}
y_{n}=b_0+b_1\left(x_{n}-b_2\right)^ 2 ,
\end{equation}
where, $b_0$ and $b_2$ are the vertical and horizontal distances of the centre of the parabola, while $b_1$ is its curvature.
The values of $b_1$, $b_2$ and the angle to orient the linear brightenings/darkenings on the vertical direction were stored in the header of the raw files, while $b_0$ was not stored as it has a different value for each arc. 
Furthermore, we make the assumption that all arcs can be described with the same parabola with different offsets in the vertical direction.
This is justified, unless there are other distortions of the image affecting the shape of the artefacts. 
We also note that our processing can return consistent results with small deviations in the determination of the curvature.
The angle of the linear segments is used at the beginning of the calibration process to rotate the image so that the linear brightenings/darkenings are on the vertical direction and the arcs are on the horizontal direction. 
The image is temporarily re-sampled prior to performing the polynomial fits in the horizontal direction, so that the arcs are straightened (see 2nd column in Fig. \ref{fig:kyotoexampleprocessing}). 
The re-sampling is done by applying a transformation of $y_{ij}=-b_1(x_{ij}-b_2)^2$,
where $y_{ij}$ and $x_{ij}$ is the distance in the vertical and horizontal direction of the $i$, $j$ pixel, respectively.
After each fit in the horizontal direction, we apply the inverse transformation on the result of the fit, so to reintroduce the curvature of the arcs. Hence we apply the polynomial fits across linear segments, but due to the aforementioned transformation the result follows the curvature of the arcs. 
The image resulting after the first transformation is somewhat egg-shaped (see 2nd column of Fig. \ref{fig:kyotoexampleprocessing}), while the circular shape of the disc is recovered after the inverse transformation.
Both the direct and its inverse transformations were applied every time a polynomial fit on the horizontal direction had to be performed. We note that this transformation affects only the result of the fits on the horizontal direction, the original image remained unchanged by this procedure. 
Besides, we adapted the width of the running window median filter to be $R/20$ instead of $R/6$ which was used for the processing of data from all the other datasets.
Examples of the processing applied to Ky data can be seen in Fig. \ref{fig:kyotoexampleprocessing}, showing that our processing could accurately remove the severe artefacts affecting the raw data. 
More such examples are shown in \cite{chatzistergos_historical_2020}.

\section{Assignment of datasets in the backbone series}
\label{sec:bakcboneasignemnt}
\begin{figure*}
	\centering
	\includegraphics[width=1\linewidth]{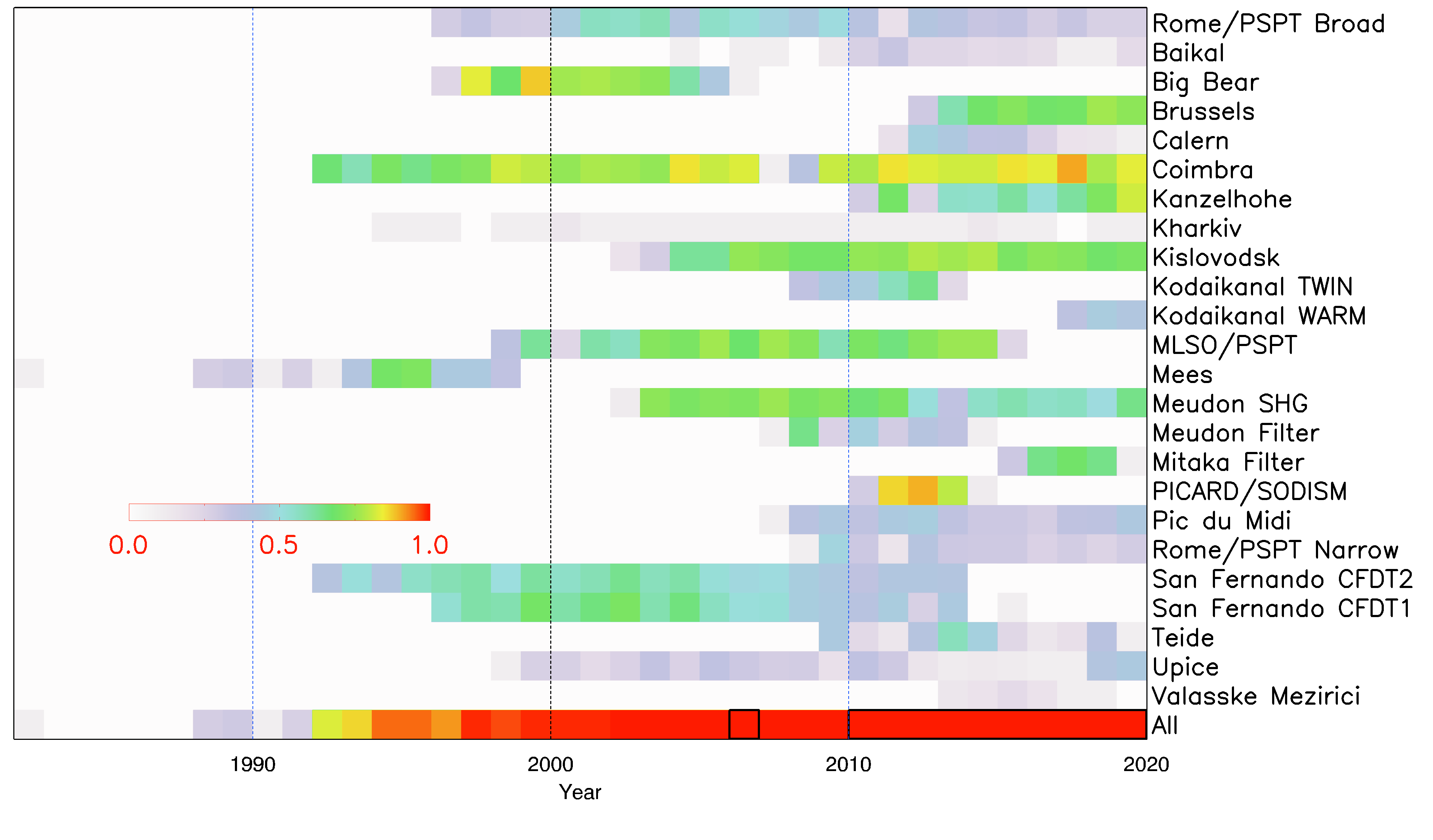}
	\caption{Annual coverage of datasets included in the RP1 backbone.}
	\label{fig:timelinepspt}
\end{figure*}

\begin{figure*}
	\centering
	\includegraphics[width=1\linewidth]{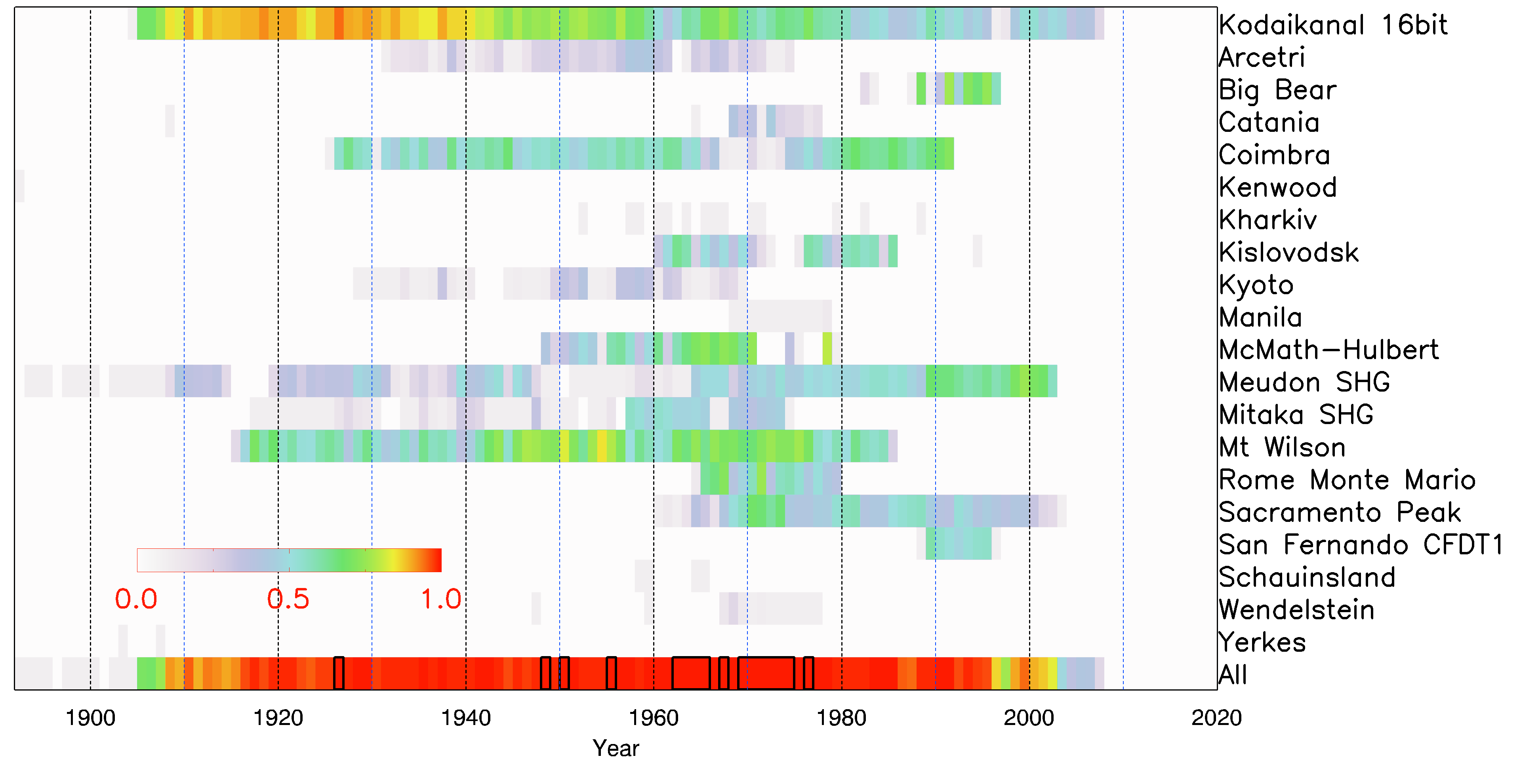}
	\caption{Annual coverage of datasets included in the Ko backbone.}
	\label{fig:timelinekoda}
\end{figure*}

\begin{figure*}
	\centering
	\includegraphics[width=1\linewidth]{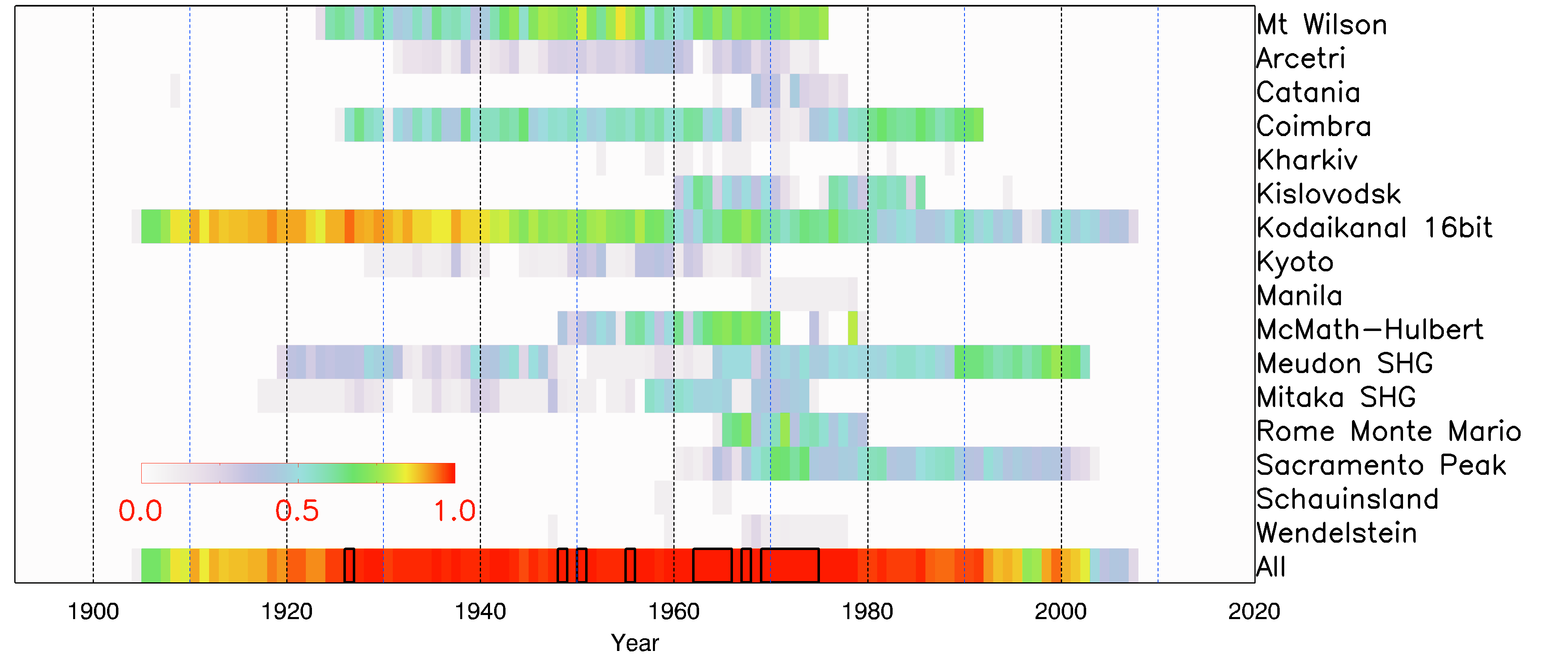}
	\caption{Annual coverage of datasets included in the MW backbone.}
	\label{fig:timelinemw}
\end{figure*}

Table \ref{tab:backbones} lists the various datasets used in this study and the backbone that they were assigned to.
The annual coverage of the datasets in the individual backbone series is shown in Fig. \ref{fig:timelinepspt}--\ref{fig:timelinemw}.

RP1 is the overall reference series for our composite.
All series with a sufficient direct overlap to RP1 were assigned to that backbone. 
This is not the case for the series from Ko and SP, as well as the early periods of BB, Co, Kh, MD1, and SF1 which were not included in the RP1 backbone due to insufficient overlap.

Following Paper 2 we split the Ar series at 25/05/1953, the Mi series at 01/02/1966, the MW series at 21/08/1923 and 01/01/1976, the Kh series at 01/01/1994, and the SP series at 01/01/1963 to account for instrumental changes affecting the coherence of the series.
Similarly, we split the BB series at 01/07/1992 and 10/09/1996 \citep{naqvi_big_2010}, the Ka series at 24/11/2012, the SF1 series at 05/02/2002, and the SF2 series at 28/07/1998 when the employed filters were replaced \citep{chapman_facular_2011}. 
We also split the ML series at 01/01/2005, because prior to that date there were many instrumental issues as evidenced by the log of the observations\footnote{\url{http://lasp.colorado.edu/media/projects/pspt_access/PSPT_Final_Release_Notes.pdf}}  \citep[see also][]{vogler_defects_2008}. 
Due to the inhomogeneities of the ML dataset, results from the data prior to 2005 have not been presented before this study.
We also split the Co series at 01/01/1992 to account for the change of the grating system and 01/01/2008 when a CCD camera started being used for the regular observations.
Similarly, we split the MD1 series at 01/01/1919, 17/09/2002, and 15/06/2017 to account for instrumental changes, the introduction of a CCD camera, and the change in bandwidth used for the observations, respectively. 
Finally, we split the Ki series at 2002 to account for the introduction of the CCD.

We merged the series by Ke\footnote{The data were downloaded from \url{http://peabody.yale.edu/collections}} and YR\footnote{The data were downloaded from \url{http://photoarchive.lib.uchicago.edu}} due to the low number of recovered images from these datasets and the fact that they were performed with the same telescope and spectroheliograph \citep{hale_rumford_1903}. 
Similarly we merged the series by Sc and WS. We will refer to these two series as Ke/YR and WS/Sc, respectively.


\begin{table*}
	\caption{List of observatories within each backbone.}        
	\label{tab:backbones}    
	\centering                  
	\begin{tabular}{l*{2}{c}}    
		\hline\hline                   
		Observatory & \multicolumn{2}{c}{Backbone}\\
							 & Our series & Individual\\
		\hline
		Ar	(before 25/05/1953)	  		&Ko, MW	& Ky, MD1, MM\\
		Ar	(after 25/05/1953)			&Ko, MW & Ar, Ky, MD1, MM, Ro, SP\\
		Ba							  	&RP1	&\\
		BB (before 01/07/1992)	  		&Ko     & MD1, SP\\
		BB (01/07/1992--10/09/1996) 	&Ko		& MD1, SP\\
		BB (after 10/09/1996) 			&RP1	&\\
		Br   			 		      	&RP1	&\\
		CT				 				&Ko, MW &Ar, MD1, MM, Ro, SP\\
		CL				 				&RP1	&\\
		Co	(before 01/01/1992)   		&Ko, MW &Ar, Ky, MD1, MM, Ro, SP\\
		Co (01/01/1992--01/01/2008) 	&RP1	&\\
		Co (after 01/01/2008) 			&RP1	&\\
		Ka	(before 24/11/2012)			&RP1	&\\
		Ka	(after 24/11/2012)  		&RP1	&\\
		Ke/YR				 			&Ko		& \\
		Kh (before 01/01/1994)			&Ko, MW	&Ar, Ky, MM, Ro \\
		Kh (after 01/01/1994)			&RP1	&\\
		Ki (before 01/01/2002)			&Ko, MW	&Ar, Ky, MD1, MM, Ro, SP\\
		Ki (after 01/01/2002)			&RP1	&\\
		Ko			 					&Ko, MW	&Ar, Ky, MD1, MM, Ro, SP\\
		KT			 					&RP1	&\\
		KW			 					&RP1	&\\
		Ky				 				&Ko, MW	&Ar, Ky, MD1, MM, Ro, SP\\
		Ma		 						&Ko, MW	&Ar, Ky, MD1, MM, Ro, SP\\
		ML (before 01/01/2005)   		&RP1	&\\
		ML (after 01/01/2005)   		&RP1	&\\
		MM		 						&Ko, MW	&Ar, Ky, MD1, MM, Ro, SP\\
		MS	   			 				&RP1 	&\\
		MD1 (before 01/01/1919)			&Ko		& \\
		MD1 (01/01/1919--26/09/2002)	&Ko, MW	&Ar, Ky, MD1, MM, Ro, SP\\
		MD1 (27/09/2002--14/06/2017)	&RP1	&\\
		MD1 (after 15/06/2017)			&RP1	&\\
		MD2	   			 				&RP1	&\\
		Mi1 (before 01/02/1966)   		&Ko, MW	&Ar, Ky, MD1, MM, Ro, SP\\
		Mi1 (after 01/02/1966)   		&Ko, MW	&Ar, Ky, MD1, MM, Ro, SP\\
		Mi2			 					&RP1	&\\
		MW (before 21/08/1923)	 		&Ko		&MD1\\
		MW (21/08/1923--31/12/1975)	 	&Ko, MW	&Ar, Ky, MD1, MM, Ro, SP\\
		MW (after 31/12/1975)	 		&Ko		&MD1, MM, Ro, SP\\
		PM			 					&RP1	&\\
		PS				 				&RP1	&\\
		Ro	 							&Ko, MW	&Ar, Ky, MD1, MM, Ro, SP\\
		RP1		     					&RP1	&\\
		RP2		     					&RP1	&\\
		SF1 (before 05/02/1996)	 		&Ko		&MD1, SP\\
		SF1 (after 05/02/1996)			&RP1	&\\
		SF2 (before 28/07/1998) 		&RP1	&\\
		SF2 (after 28/07/1998) 			&RP1	&\\
		SP (before 01/01/1963)			&Ko, MW	&Ar, Ky, MD1, MM\\
		SP (after 01/01/1963)			&Ko, MW	&Ar, Ky, MD1, MM, Ro, SP\\
		Te      			 			&RP1	&\\
		UP      			 			&RP1	&\\
		VM 								&RP1	&\\
		WS/Sc  		 					&Ko, MW	&Ar, Ky, MD1, MM, Ro, SP\\
		\hline
	\end{tabular}
	\tablefoot{Columns are: name of the observatory and backbone to which it has been assigned in our composite series as well as the individual backbone series discussed in Sect. \ref{sec:discussion}.}
\end{table*}

\end{document}